\RequirePackage[2020-02-02]{latexrelease}
\documentclass[twocolumn]{aastex631}
\usepackage{hyperref}
\hypersetup{
  colorlinks=true,
  allcolors=cyan
}

\usepackage{amsmath, mathtools, amsthm, amssymb}
\usepackage{enumitem}
\usepackage[toc,page]{appendix}
\usepackage{soul}
\usepackage{graphicx}
\usepackage{appendix}
\graphicspath{{./}}
\usepackage{xcolor}
\usepackage[super]{nth}

\newcommand\soutm{\bgroup\markoverwith
{\textcolor{black}{\rule[0.5ex]{2pt}{0.8pt}}}\ULon}

\shorttitle{AGN feedback prescription}
\shortauthors{Contini et al.}

\begin{document}

\title{A Full AGN Feedback Prescription for Numerical Models: Negative, Positive and Hot Gas-Ejection Modes}

\author{Emanuele Contini}
\affil{Department of Astronomy and Yonsei University Observatory, Yonsei University, 50 Yonsei-ro, Seodaemun-gu, Seoul 03722, Republic of Korea}
\email{emanuele.contini82@gmail.com}
\author{Sukyoung K. Yi}
\affil{Department of Astronomy and Yonsei University Observatory, Yonsei University, 50 Yonsei-ro, Seodaemun-gu, Seoul 03722, Republic of Korea}
\email{yi@yonsei.ac.kr}
\author{Jinsu Rhee}
\affil{Department of Astronomy and Yonsei University Observatory, Yonsei University, 50 Yonsei-ro, Seodaemun-gu, Seoul 03722, Republic of Korea}
\affil{Korea Astronomy and Space Science Institute, 776, Daedeokdae-ro, Yuseong-gu, Daejeon 34055, Republic of Korea}
\author{Seyoung Jeon}
\affil{Department of Astronomy and Yonsei University Observatory, Yonsei University, 50 Yonsei-ro, Seodaemun-gu, Seoul 03722, Republic of Korea}

\begin{abstract}
We build upon the state-of-the-art semi-analytic model \texttt{FEGA24} (Formation and Evolution of GAlaxies, \citealt{contini2024d}), which integrates the latest prescriptions relevant to galaxy formation and evolution, alongside a comprehensive AGN feedback model. This model incorporates three modes of feedback: negative (preventing excessive cooling), positive (enhancing star formation), and hot gas ejection (expelling gas beyond the virial radius of halos). These modes operate in a coordinated manner: the negative mode regulates the cooling process, the positive mode promotes bursts of star formation, and the hot gas ejection mode expels gas beyond the virial radius when the AGN is sufficiently powerful.
Our updated semi-analytic model, \texttt{FEGA25}, retains the qualitative and quantitative consistency of the analyses presented in \cite{contini2024d}, while delivering more robust results. Notably, \texttt{FEGA25} provides a more detailed characterization of the fraction of passive galaxies as a function of stellar mass and redshift, predicts a main sequence of star-forming galaxies more consistent with observations and a more accurate cosmic star formation rate density with redshift. Moreover, it estimates the fraction of hot gas in halos closer to observed values.
These findings underscore the importance of a physical mechanism capable of ejecting hot gas beyond the virialized region of dark matter halos without significantly altering the stellar and cold gas components. Such a mechanism is crucial to ensure the proper functioning of other processes, such as cooling and star formation. Since supernova feedback is already modeled at its maximum efficiency, AGN feedback emerges as the natural candidate for this role.
\end{abstract}


%%%%%%%%%%%%%%%%%%%%%%%%%%%%%%%%%%%%%%%%%%%%%%%%%%%%%%%%%%%%%%%%%%%%%%%%%%%%%%%%%%%%%%%%%%%%%%%%%%%%%%%%%%%%%%%%%%%%%%%%%%%%%%%%%%%%%%%%%%
\section{Introduction}
\label{sec:intro}

In the intricate realm of galaxy formation and evolution, numerous physical processes interact and converge to shape the diverse range of galaxies observed, with their properties varying according to environment, size, and epoch (\citealt{kauffmann2003,baldry2004,delucia2012,rhee2020,henriques2017,contini2020,jeon2022,delucia2024} and many others). Numerical methods have continuously evolved to better capture these complexities, incrementally assembling pieces of a larger puzzle (\citealt{rantala2017,shankar2017,moster2018,cora2018,kravtsov2018,behroozi2019,contini2020,henriques2020,springel2021,dubois2021,pakmor2023,stevens2024,contini2024d}, among others).

From a semianalytic perspective, these models offer a computationally efficient approach compared to fully hydrodynamic simulations. Their relatively low cost allows for isolating specific processes to assess their effects on global scales. Over the years, semianalytic models (hereafter SAMs) have proven to be robust tools for studying galaxy formation and evolution. Recent advancements have significantly improved their ability to describe and predict galaxy properties (\citealt{guo2011,guo2013,contini2014,henriques2015,hirschmann2016,cora2018,xie2020,henriques2020,delucia2024,stevens2024,contini2024d}). By leveraging merger trees derived from numerical simulations, SAMs incorporate the physics of baryons to reproduce the observed diversity of galaxies. However, despite their successes, even the most sophisticated SAMs face challenges, often excelling in some areas while falling short in others.

One persistent issue in SAMs and in state-of-the-art numerical simulations is the overestimation of the hot gas fraction within halos. While predictions align well with observations at cluster scales (e.g., \citealt{cui2022,angelinelli2023,robson2023}), they significantly exceed observed values at smaller halo masses (\citealt{popesso2024,dev2024}, and references therein). Many contemporary simulations, such as Illustris (\citealt{genel2014}), Bahamas (\citealt{mccarthy2017}), IllustrisTNG (\citealt{pillepich2018}), Simba (\citealt{dave2019}), MillenniumTNG (\citealt{pakmor2023}), and Flamengo (\citealt{schaye2023}), still exhibit similar discrepancies. A recent study by \cite{popesso2024}, which compared eROSITA eFEDS field data (\citealt{brunner2022}) to these simulations, revealed a systematic overprediction of the hot gas fraction across a wide range of halo masses—except for Illustris, which underpredicts it. Interestingly, the Magneticum simulation (\citealt{dolag2016}) is the only one consistent with observations. These findings suggest that a key physical mechanism capable of relocating hot gas beyond the virial radius—without acting on the cold gas and stellar components—may be absent in most models.

Such a mechanism is hypothesized to stem from a non-gravitational process capable of altering the thermodynamic properties of hot gas while leaving the rest of the baryonic content in galaxy groups—and to a lesser extent, in galaxy clusters where the discrepancy between models and observations is less pronounced—essentially unaffected. Supernova feedback is one possible candidate, potentially effective in low-mass galaxy groups. However, in most numerical models, including semianalytic approaches, this feedback is already implemented at near-maximal efficiency (see, e.g., \citealt{guo2013,henriques2015,hirschmann2016,delucia2024,contini2024d}).

Another viable mechanism, particularly relevant in larger groups and extending to galaxy clusters, is feedback driven by active galactic nuclei (AGN). The energy injected by black holes (BHs) in many scenarios can be extraordinarily high (e.g., \citealt{fabjan2010,lebrun2014,biffi2018,eckert2021,popesso2024} and references therein), making AGN feedback a potentially significant factor in influencing the hot gas content, even at larger scales in smaller groups (see \citealt{oppenheimer2021}).

In \citealt{contini2024d} (hereafter C24), we introduced {\small FEGA} (Formation and Evolution of GAlaxies), a SAM incorporating state-of-the-art prescriptions for galaxy formation processes, including a novel implementation of the observed positive AGN feedback (e.g., \citealt{zinn2013,salome2015,cresci2015a,salome2017,shin2019,nesvadba2020,joseph2022,venturi2023,gim2024}). Motivated by the need to address the persistent overestimation of hot gas within virialized regions of dark matter (DM) halos, this work presents a new version of the model, \texttt{FEGA25}, which explicitly accounts for gas ejection beyond the virial radius.

The updated \texttt{FEGA25} includes a more physically grounded implementation of positive AGN feedback, now integrated with the negative mode, thereby reducing the number of free parameters. Additionally, it introduces a third mode of AGN feedback, designed to eject hot gas beyond the halo. These modes are unified under a single parameter governing the BH accretion efficiency. This new version also features a detailed implementation of stellar halos for central galaxies (\citealt{contini2024e}) and aims to bring the predicted hot gas fraction closer to observed values. As with the previous version, \texttt{FEGA25} is expected to deliver improvements in other galaxy properties and scaling relations as well.

In Section~\ref{sec:model}, we describe the three AGN feedback modes (radio mode) implemented in the model and provide a brief overview of the calibration process. Section~\ref{sec:results} presents a detailed analysis of the key advancements introduced in \texttt{FEGA25}, including refined star formation rate and passive fraction–stellar mass relations, as well as a reduction in the predicted hot gas fraction as a function of halo mass. We also explore the impact of the updated positive AGN feedback implementation compared to the previous version. This section discusses also the broader implications of our findings, while Section~\ref{sec:conclusion} summarizes the main conclusions of this study. Throughout this work, we adopt a Chabrier (\citealt{chabrier2003}) initial mass function for stellar mass calculations, and all units are h-corrected.

%%%%%%%%%%%%%%%%%%%%%%%%%%%%%%%%%%%%%%%%%%%%%%%%%%%%%%%%%%%%%%%%%%%%%%%%%%%%%%%%%%%%%%%%%%%%%%%%%%%%%%%%%%%%%%%%%%%%%%%%%%%%%%%%%%%%%%%%%%%
\section[]{FEGA25}
\label{sec:model}

\texttt{FEGA25} is an updated version of the earlier model presented in C24 (version ModA), hereafter \texttt{FEGA24}, which incorporates state-of-the-art prescriptions for the key processes governing galaxy formation and evolution. \texttt{FEGA24} features two significant innovations: (1) a more realistic representation of star formation based on the extended Kennicutt-Schmidt relation (\citealt{kennicutt1998,shi2018}); and (2) an unprecedented modeling of positive AGN feedback. This SAM has demonstrated remarkable accuracy in reproducing key galaxy scaling relations and properties, ranging from the evolution of the stellar mass function (SMF, with which it was calibrated) to genuine predictions of star formation rates, specific star formation rates, galaxy morphologies, and scaling relations, including stellar-to-halo mass, metallicity-mass, red fraction-mass, and BH mass versus bulge/galaxy mass relations.

Despite its reliability, \texttt{FEGA24} revealed several areas for potential improvement. For instance, although not explicitly highlighted in C24, \texttt{FEGA24} fails to accurately predict the observed fraction of hot gas in halos, producing a systematic overestimation compared to observations. As discussed in Section~\ref{sec:intro}, this issue is not exclusive to semianalytic models but also affects numerical simulations (see, e.g., \citealt{popesso2024}). Addressing this challenge necessitates introducing a physical mechanism to reduce the excess hot gas. We propose such a mechanism in the form of a third mode of AGN feedback, capable of ejecting hot gas beyond the virial radius without altering the stellar and cold gas components.

\texttt{FEGA25} thus represents a more advanced version of \texttt{FEGA24}, incorporating three key features: (a) a more physical treatment of positive AGN feedback, closely linked to the negative mode; (b) an efficient mechanism for expelling gas beyond the virial radius; and (c) a comprehensive description of stellar halo formation, characterizing these regions as transitional zones between central galaxies and their associated diffuse light (\citealt{contini2024e}).

In the remainder of this Section, we detail the three modes of AGN feedback implemented in \texttt{FEGA25} and provide a brief overview of the calibration method employed for the model.

%%%%%%%%%%%%%%%%%%%%%%%%%%%%%%%%%%%%%%%%%%%%%%%%%%%%%%%%%%%%%%%%%%%%%%%%%%%%%%%%%%%%%%%%%%%%%%%%%%%%%%%%%%%%%%%%%%%%%%%%%%%%%%%%
\subsection[]{AGN Feedback: Negative, Positive and Ejection Modes}
\label{sec:agnfeedback}

The AGN feedback plays a crucial role in galaxy formation, primarily by regulating the accretion of hot gas onto the BH and controlling the growth of the host galaxy. In the standard framework, AGN feedback operates in two main modes: (1) the quasar mode, responsible for BH growth during galaxy mergers; and (2) the radio mode, which is believed to limit excessive galaxy growth by injecting energy to counteract ongoing cooling. The energy injected is proportional to the accretion of hot gas onto the BH and is governed by the efficiency of this process. For an overview of the quasar mode in \texttt{FEGA24}, we refer the reader to C24, while a more detailed description can be found in the original implementation by \cite{croton2006}. Here, it is worth mentioning that BHs naturally form during the quasar mode through the accretion of cold gas, in an amount defined by Equation 20 in C24. They subsequently grow through both the radio mode (hot gas accretion) and the quasar mode (cold gas accretion and mergers).

%%%%%%%%%%%%%%%%%%%%%%%%%%%%%%%%%%%%%%%%%%%%%%%%%%%%%%%%%%%%%%%%%%%%%%%%%%%%%%%%%%%%%%%%%%%%%%%%%%%%%%%%%%%%%%%%%%%%%%%%%%%%%%%%
\subsubsection[]{Negative Feedback}
\label{sec:negativefb}

AGN feedback is commonly understood as a mechanism that prevents gas from cooling, functioning as a form of negative feedback that counteracts the formation of new stars. In the radio mode, the injection of energy capable of halting gas cooling is implemented as described in the original version by \cite{croton2006}. The central BH accretes a portion of the available hot gas, with the accretion rate defined as
\begin{equation}\label{eqn:radiomode}
\dot{M}_{\rm{BH}} = \kappa_{\rm{AGN}} \left(\frac{f_{\rm{hot}}}{0.1}\right)\left(\frac{V_{200}}{200\,  \rm{km/s}}\right)^3 \left(\frac{M_{\rm{BH}}}{10^8 \, M_{\odot}/h}\right)
\end{equation}
expressed in units of $M_{\odot}/\rm{yr}$. Here, $f_{\rm{hot}}$ represents the ratio of hot gas mass to dark matter mass, $\kappa_{\rm{AGN}}$ is a parameter controlling the accretion efficiency, $V_{200}$ is the virial velocity, and $M_{\rm{BH}}$ is the BH mass. The parameter $\kappa_{\rm{AGN}}$ is critical in this mode, as well as in subsequent modes, as it governs gas accretion and the energy release into the surrounding hot gas. Following C24, this parameter is calibrated during the model setup.

The mechanical energy released by the BH during gas accretion, as described by Equation~\ref{eqn:radiomode}, is given by
\begin{equation}\label{eqn:energy}
\dot{E}_{\rm{radio}} = \eta_{\rm{rad}} \dot{M}_{\rm{BH}}c^2 ,
\end{equation}
where $\eta_{\rm{rad}}$ is a parameter set to 0.1, and $c$ is the speed of light. From Equations~\ref{eqn:radiomode} and~\ref{eqn:energy}, the net rate of gas cooling is expressed as
\begin{equation}\label{eqn:mcool}
\dot{M}_{\rm{cool,new}} = \dot{M}_{\rm{cool}}-2\frac{\dot{E}_{\rm{radio}}}{V_{200}^2},
\end{equation}
where $\dot{M}_{\rm{cool}}$ is the original cooling flow.

The injection of mechanical energy described above can entirely halt gas cooling, by reducing $\dot{M}_{\rm{cool,new}}$ to zero. However, this mode of AGN feedback operates with varying efficiency over time, being almost negligible at high redshifts and becoming increasingly significant at lower redshifts, favored by the increasing availability of hot gas, partly resulting from the reincorporation of gas previously ejected beyond the halos' virial radius. Moreover, the efficiency of this negative mode depends on halo mass, being more effective in halos with $\log M_{\rm{halo}} > 13$.

As highlighted in C24, Equation~\ref{eqn:mcool} defines the net cooling gas rate in a generic SAM, marking the endpoint of AGN feedback. However, \texttt{FEGA24} also incorporates a second mode—positive feedback that enhances star formation—a feature retained in \texttt{FEGA25}.

%%%%%%%%%%%%%%%%%%%%%%%%%%%%%%%%%%%%%%%%%%%%%%%%%%%%%%%%%%%%%%%%%%%%%%%%%%%%%%%%%%%%%%%%%%%%%%%%%%%%%%%%%%%%%%%%%%%%%%%%%%%%%%%%
\subsubsection[]{Positive Feedback}
\label{sec:positivefb}

The positive AGN feedback, first introduced in C24, was an innovative attempt to incorporate a physical mechanism into SAMs, supported by various observational studies (\citealt{zinn2013,salome2015,cresci2015a,cresci2015b,mahoro2017,salome2017,shin2019,nesvadba2020,joseph2022,tamhane2022,venturi2023,gim2024}) as well as other theoretical and numerical approaches (e.g., \citealt{gaibler2012,zubovas2013,silk2013,bieri2015,bieri2016,zubovas2017,mukherjee2018,silk2024}), where the AGN can trigger star formation by induced pressure at the edge of the outflows (\citealt{cresci2015a}), or enhance star formation via burst in different locations depending on the power of the feedback (e.g., \citealt{silk2013}). However, as noted in the previous work, the prescription used was more an attempt to isolate the positive feedback effect rather than a fully physical model. The goal was to demonstrate that both the negative and positive modes of AGN feedback can coexist without altering the overall properties of galaxies and their scaling relations.

In C24, the positive feedback mode was modeled by assuming a burst of star formation after the negative feedback had finished its effect. Specifically, we proposed that part of the cooling gas in Equation~\ref{eqn:mcool} could form stars, but only if the energy injected by the BH was insufficient to completely balance the cooling gas, ensuring that $\dot{M}_{\rm{cool,new}} > 0$. We described this burst of star formation due to positive feedback with the following functional form:
\begin{equation}\label{eqn:pAGN}
\dot{M}_* = \alpha_{\rm{pAGN}}\left(\frac{M_{\rm{BH}}}{10^{8} \, M_{\odot}/h}\right)^{\beta_{\rm{pAGN}}}\dot{M}_{\rm{cool,new}},
\end{equation}
where $\dot{M}_*$ represents the star formation rate, which depends on $\dot{M}_{\rm{cool,new}}$ and is proportional to the BH mass. The parameters $\alpha_{\rm{pAGN}}$ and $\beta_{\rm{pAGN}}$ determine the fraction of cooling gas converted into stars, and were calibrated during the model setup.

In \texttt{FEGA25}, we re-calibrate the model so that both the negative and positive feedback modes are described by the same parameters, meaning that no additional free parameters are introduced for the positive mode. We also maintain the assumption from \texttt{FEGA24} that the positive mode only activates when the negative mode has not fully suppressed cooling ($\dot{M}_{\rm{cool,new}} > 0$). Following a similar approach, the new prescription for the burst of star formation is:
\begin{equation}\label{eqn:newpAGN}
\dot{M}_* = \left(\frac{\delta M_{\rm{BH}}}{M_{\rm{BH}}}\right) \left(\frac{\dot{M}_{\rm{cool,new}}}{10^{8}M_{\odot}/h}\right),
\end{equation}
where $\delta M_{\rm{BH}}$ represents the net increase in BH mass from hot gas accretion, as described by Equation~\ref{eqn:radiomode}.

Compared to Equation~\ref{eqn:pAGN}, Equation~\ref{eqn:newpAGN} provides a more physically-motivated implementation of positive AGN feedback for several reasons. The most significant is that the efficiency of this mode now depends directly on the efficiency of the negative mode through the term $\frac{\delta M_{\rm{BH}}}{M_{\rm{BH}}} \propto \kappa_{\rm{AGN}}$, which incorporates the accretion efficiency onto the BH as governed by $\kappa_{\rm{AGN}}$. This way, the efficiencies of both feedback modes are controlled by a single parameter, $\kappa_{\rm{AGN}}$, enabling the MCMC algorithm to focus on calibrating this parameter. Another important aspect of this approach is that it links the two modes: the positive mode can be strong when the negative mode is weak, and vice versa, or they can both be strong simultaneously. A key expectation is that the positive mode will dominate at high redshifts, when the negative mode is less effective, and will fade at lower redshifts, when the negative mode is more active.

%%%%%%%%%%%%%%%%%%%%%%%%%%%%%%%%%%%%%%%%%%%%%%%%%%%%%%%%%%%%%%%%%%%%%%%%%%%%%%%%%%%%%%%%%%%%%%%%%%%%%%%%%%%%%%%%%%%%%%%%%%%%%%%%
\subsubsection[]{Hot Gas Ejection}
\label{sec:ejection}

In Section~\ref{sec:intro}, we briefly discussed a significant challenge in SAMs: their inability to predict the correct fraction of hot gas in halos, along with a potential solution. This solution needs to involve a mechanism that can expel some of the hot gas beyond the virial radius while having no impact on cooling or star formation processes. In other words, the mechanism must remove hot gas without affecting the stellar or cold gas components. To address this, we introduce a new mode of AGN feedback in \texttt{FEGA25}: the hot gas ejection mode.

As previously mentioned, the negative AGN feedback can sometimes be powerful enough to fully counterbalance the cooling flow, meaning that the injected energy is sufficient to heat the gas back to the virial temperature. But what if this energy is also enough to push gas outside the virialized region? One possible explanation is that in situations where the energy injected by the AGN exceeds what is needed to halt the cooling flow, the excess energy could push gas outside the halo. Therefore, the difference between the energy used to heat the gas and the energy responsible for cooling could be assumed to expel gas beyond the virial radius.

Following a similar approach to that used in C24 for isolating the positive feedback, we assume that such a mechanism is feasible,
and propose that the AGN can expel a certain amount of hot gas and deposit it into an "ejected" reservoir. The SAM incorporates this by assuming that the amount of hot gas expelled depends on the efficiency of the other two AGN feedback modes ($\kappa_{\rm{AGN}}$) and is directly linked to the virial velocity of the host halo. Consequently, it becomes more difficult to expel gas from larger halos. The following equation in \texttt{FEGA25} computes the mass of hot gas that is ejected:
\begin{equation}\label{eqn:ejection}
M_{\rm{ejected}} = \left(\frac{\delta M_{\rm{BH}}}{M_{\rm{BH}}}\right) \left(\frac{M_{\rm{hot}}}{10^{8}M_{\odot}/h}\right)\left(1-\frac{V_{\rm{200}}}{V_{\rm{scale}}}\right),
\end{equation}
where $V_{\rm{scale}}$ is a free parameter, calibrated during the model fitting process, that controls the efficiency threshold. \footnote{This prescription is applied even when $\dot{M}_{\rm{cool}}>2\frac{\dot{E}_{\rm{radio}}}{V_{200}^2}$, i.e., when the injection of energy is not sufficient to completely stop the cooling of gas.}
Larger values of $V_{\rm{scale}}$ enhance the efficiency in larger halos, while smaller values reduce the efficiency in lower-mass halos. The ejected gas mass, $M_{\rm{ejected}}$, is then stored in the ejected reservoir, but it remains available for later reincorporation, as described in C24. Clearly, if $V_{200}>V_{\rm{scale}}$, the amount of hot gas ejected is assumed to be 0.

In conclusion, the combination of negative, positive, and hot gas ejection modes describes the full AGN feedback implemented in \texttt{FEGA25}. This version differs from \texttt{FEGA24} in C24 by providing a more physically grounded treatment of the positive feedback mode (see Equation~\ref{eqn:newpAGN}) and introducing the new hot gas ejection mode (Equation~\ref{eqn:ejection}). As a result, \texttt{FEGA25} is the only SAM that includes both a prescription for positive feedback and a mechanism for hot gas ejection. It is clear that this update must improve the model's ability to describe galaxy properties and their scaling relations, while also addressing the issue of excess hot gas in halos, without compromising the other properties and relations of galaxies.

%%%%%%%%%%%%%%%%%%%%%%%%%%%%%%%%%%%%%%%%%%%%%%%%%%%%%%%%%%%%%%%%%%%%%%%%%%%%%%%%%%%%%%%%%%%%%%%%%%%%%%%%%%%%%%%%%%%%%%%%%%%%%%%
\subsection[]{Calibration}
\label{sec:calibration}

The calibration of a SAM is a crucial step, especially when the prescriptions used to describe the various physical processes involved in galaxy formation include numerous parameters that are often poorly constrained by observations or even entirely unknown. As mentioned earlier, we calibrate \texttt{FEGA25} in the same way we calibrated \texttt{FEGA24} in C24, that is, using an MCMC approach, which is a powerful tool given its ability to reproduce a wide range of galaxy properties and scaling relations not used in the calibration. For further details on this approach, including a discussion of its advantages and disadvantages, as well as a review of alternative methods, we refer the reader to C24, \cite{henriques2020}, and the references therein.

The choice of observational data to be used for calibrating a model is critical. While a broad set of observed data for different properties or scaling relations can help better constrain the free parameters, it may also lead to biasing the model towards fitting specific data points. A reasonable compromise would be to cover as many critical properties or scaling relations as possible with observational data, providing a wide spectrum of information for the MCMC. However, this solution would leave little room for genuine predictions from the model itself. Our approach, as in C24, is more stringent: we calibrate the model solely based on the evolution of the SMF from high redshift to the present time. The SAM is then required, through calibration, to provide a good description of the SMF at each redshift, while simultaneously making genuine predictions for several galaxy properties and scaling relations. This approach has been very successful for calibrating \texttt{FEGA24}, particularly for its ModA version (see details in C24), and we built \texttt{FEGA25} by integrating the AGN feedback prescriptions described above into the former ModA from C24.

\texttt{FEGA25} is calibrated using the same set of observed SMF data used in C24, covering redshifts from $z=3$ to $z=0$. The dataset includes contributions from multiple studies to offer a comprehensive view of the SMF across various redshifts: $z=3$, $z=2.75$, $z=2$, $z=1.75$, $z=1$, $z=0.75$, $z=0.4$, and $z=0$. The SMFs for $z>0$ are taken from \cite{marchesini2009}, \cite{marchesini2010}, \cite{ilbert2010}, \cite{sanchez2012}, \cite{muzzin2013}, \cite{ilbert2013}, and \cite{tomczak2014}, while the present-day SMF is taken from \cite{baldry2008}, \cite{li-white2009}, \cite{baldry2012}, and \cite{bernardi2018}. The final component necessary for calibrating the SAM and constructing galaxy catalogs is a suite of cosmological simulations. These provide the merger trees, which serve as input for the model and contain the relevant information about the DM halos within which galaxies form.

We take advantage of the YS50HR simulation, with a cosmological volume of $(50\, \rm{Mpc/h})^3$, primarily used for the initial calibration phase, and the larger YS200 simulation, with a volume of $(200\, \rm{Mpc/h})^3$, used for the final calibration and constructing the final galaxy catalogs. These simulations cover a redshift range from $z=63$ to the present time, with 100 discrete snapshots from $z=20$ to $z=0$. The simulations are run using {\small GADGET4} (\citealt{springel2021}), with cosmological parameters consistent with the Planck 2018 cosmology (\citealt{planck2020}): $\Omega_m=0.31$ (total matter density), $\Omega_{\Lambda}=0.69$ (cosmological constant), $n_s=0.97$ (primordial spectral index), $\sigma_8=0.81$ (power spectrum normalization), and $h=0.68$ (normalized Hubble parameter). The mass resolutions of the simulations are $M_{\rm{DM,part}}=10^7 - 3.26\cdot10^8\, M_{\odot}/h$, for YS50HR and YS200, respectively.
A more detailed description of the simulation suite can be found in \cite{contini2023}.

%%%%%%%%%%%%%%%%%%%%%%%%%%%%%%%%%%%%%%%%%%%%%%%%%%%%%%%%%%%%%%%%%%%%%%%%%%%%%%%%%%%%%%%%%%%%%%%%%%%%%%%%%%%%%%%%%%%%%%%%%%%%%%%%
%%%%%%%%%%%%%%%%%%%%%%%%%%%%%%%%%%%%%%%%%%%%%%%%%%%%%%%%%%%%%%%%%%%%%%%%%%%%%%%%%%%%%
\subsubsection{Set of Parameters}\label{sec:par_set}

Building on the previous calibration of \texttt{FEGA24} ModA from C24 and the subsequent improvements made by the implementation of stellar halos in \cite{contini2024e}, we can complete the final calibration of \texttt{FEGA25}. In this process, we calibrate the relevant parameters within the AGN feedback modes, as well as the parameter $\gamma_2$ \footnote{This parameter enters in the definition of the mass of ejected gas that turns back to the hot gas available inside the virial radius, as follows: $\dot{M}_{\rm{ej}} = \gamma_2 \frac{M_{\rm{ej}}}{t_{\rm{reinc}}}$, where $M_{\rm{ej}}$ is the available gas in the ejecta and $t_{\rm{reinc}}$ is the reicorporation time. More details can be found in C24.} from C24, which plays a role in the computation of the reincorporation time of ejected gas. This parameter is particularly important due to the introduction of the hot gas ejection mode, which adds more gas to the ejected reservoir. The parameters constrained during the calibration are: the efficiency $\kappa_{\rm{AGN}}$ of the BH in Equation~\ref{eqn:radiomode}, the scale velocity $V_{\rm{scale}}$ in Equation~\ref{eqn:ejection}, and the reincorporation factor $\gamma_2$. The values obtained during calibration are the following for \texttt{FEGA25} (\texttt{FEGA24}): $\kappa_{\rm{AGN}}=2\rm{e}^{-5}\, (5\rm{e}^{-5})$ $M_{\odot}$/yr; $V_{\rm{scale}}=500$ (not applicable) km/s; $\gamma_2 =0.378\, (0.375)$.

A quick inspection of the parameter values reveals one key difference between \texttt{FEGA24} and \texttt{FEGA25}. While the reincorporation factor $\gamma_2$ remains nearly unchanged, the efficiency parameter $\kappa_{\rm{AGN}}$ in \texttt{FEGA25} is 2.5 times smaller than in \texttt{FEGA24}. The minimal change in $\gamma_2$ suggests that the continuous addition of hot gas to the ejected reservoir from the hot gas ejection mode does not significantly affect the reincorporation time of the expelled gas. On the other hand, the lower value of $\kappa_{\rm{AGN}}$ means that less hot gas is available for BH accretion, resulting in a lower energy injection required to counterbalance the cooling flow. Consequently, the BH accretes less hot gas. 

\begin{figure*}[t!]
\begin{center}
\begin{tabular}{cc}
\includegraphics[width=0.47\textwidth]{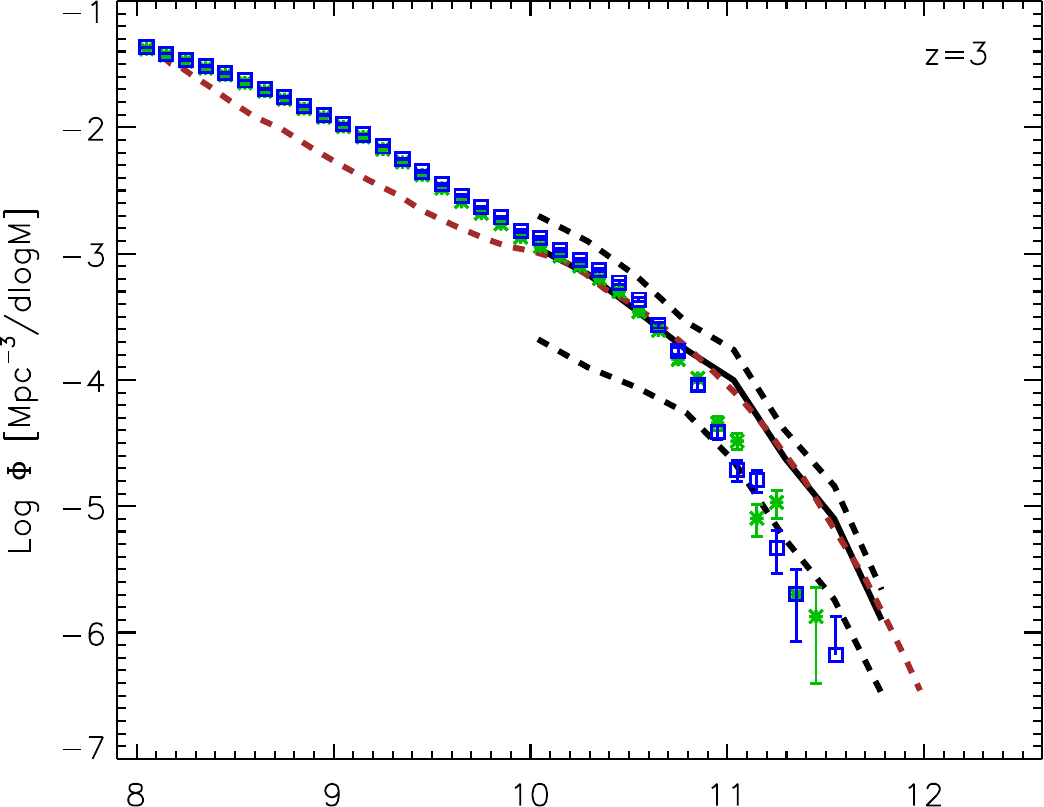} &
\includegraphics[width=0.44\textwidth]{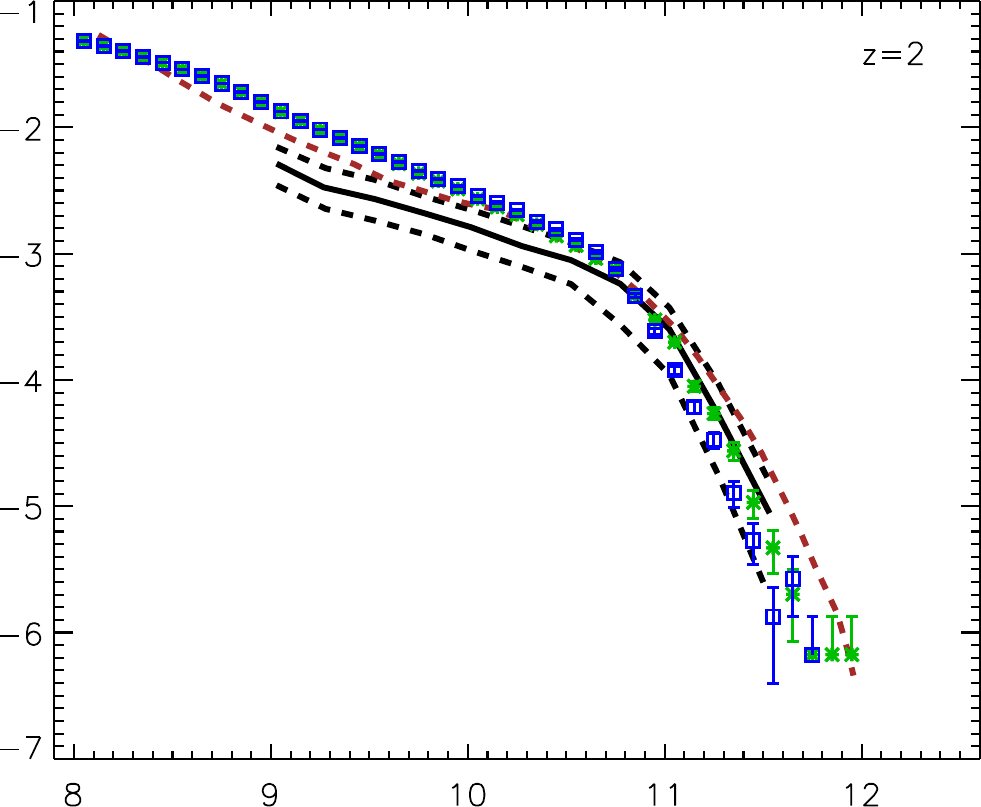} \\
\includegraphics[width=0.47\textwidth]{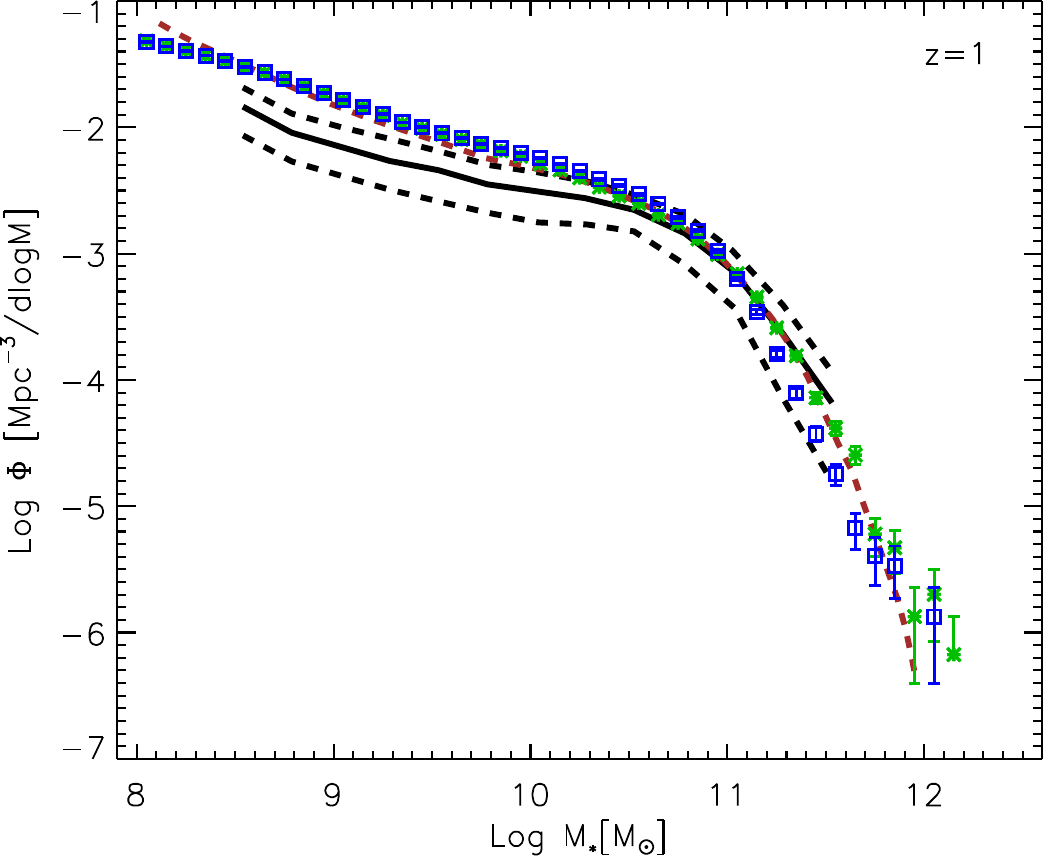} &
\includegraphics[width=0.44\textwidth]{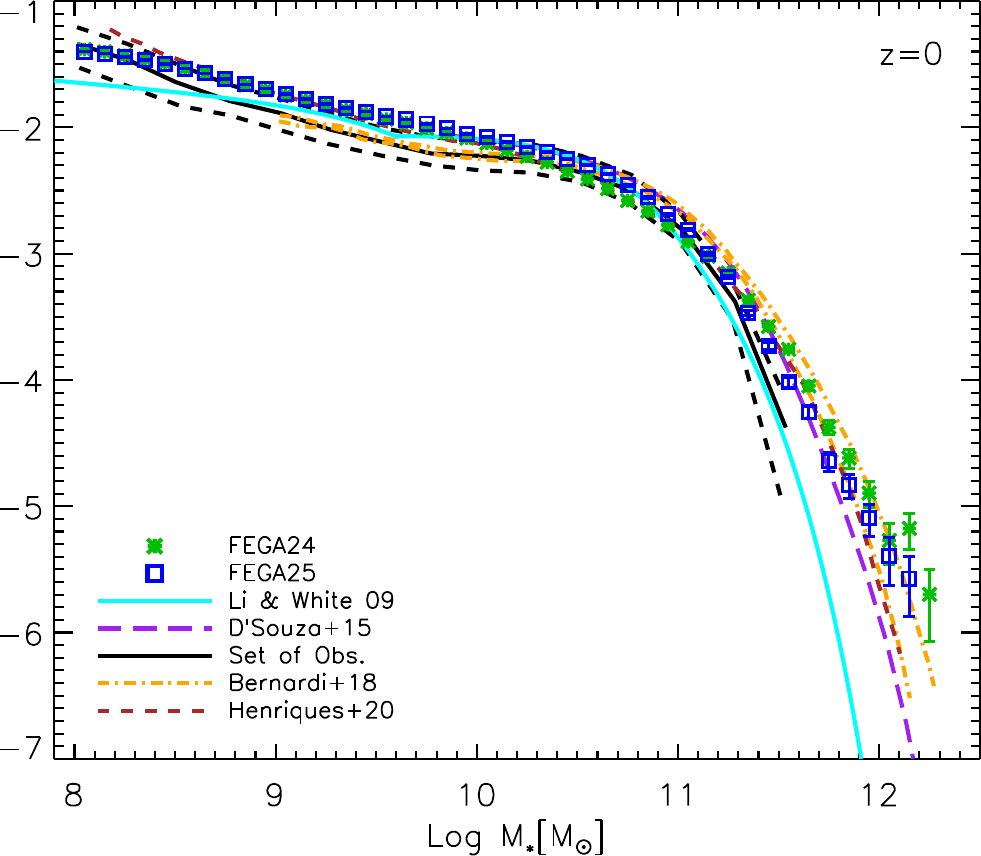}
\end{tabular}
\caption{The evolution of the SMF from $z=3$ to $z=0$ is shown, as predicted by \texttt{FEGA24} (represented by green stars) and \texttt{FEGA25} (represented by blue squares), compared with the set of observations used to calibrate the models (black lines). Additionally, we compare our predictions with those of \texttt{L-Galaxies SAM} (\citealt{henriques2020}), as well as another set of observations (at $z=0$ only) from different authors: \citealt{li-white2009} (represented by a cyan line), \citealt{dsouza2015} (represented by a purple line), and \citealt{bernardi2018} (represented by magenta lines indicating the $\pm\sigma$ region). Both models accurately describe the evolution of the SMF down to $z=0$.}
\label{fig:SMFevol}
\end{center}
\end{figure*}

Before delving into the analysis, it is essential to address a few caveats. First, during the calibration, the model was required to match the high-mass end of the SMF at $z=0$ as provided by \cite{bernardi2018}, which has been shown to offer a more accurate prediction of the stellar-to-halo mass relation at cluster scales in C24. The primary model parameter that controls the high-mass end of the SMF is the AGN efficiency, $\kappa_{\rm{AGN}}$. As a result, a higher value of $\kappa_{\rm{AGN}}$ is necessary to align the predicted SMF at $z=0$ with the observed data. However, the range of values obtained by the MCMC algorithm encompasses all possible high-mass ends (see bottom-right panel of Figure~\ref{fig:SMFevol}), with the most probable value corresponding to the desired high-mass end.

Secondly, it is important to note that the three parameters calibrated in this process are well constrained. Both $\kappa_{\rm{AGN}}$ and $\gamma_2$ remain stable across the mass range considered, while $V_{\rm{scale}}$ is more sensitive to the specific partition of the merger tree used, as different partitions sample a variety of environments. Nonetheless, $V_{\rm{scale}}$ also stabilizes once calibrated on large partitions of the YS200 simulation box.

%%%%%%%%%%%%%%%%%%%%%%%%%%%%%%%%%%%%%%%%%%%%%%%%%%%%%%%%%%%%%%%%%%%%%%%%%%%%%%%%%%%%%%%%%%%%%%%%%%%%%%%%%%%%%%%%%%
\section{Results and Discussion}
\label{sec:results}

In this section, we present our analysis focused on the key advancements introduced by the new implementations, with particular emphasis on the role of the third AGN feedback mode—namely, the hot gas ejection mode. In Section \ref{sec:smf}, we briefly describe and comment on the evolution of the SMF, which serves as the main observable for calibrating both models. Section \ref{sec:models} highlights the primary differences between the models, based on a repetition of the analysis performed in C24. Finally, in Section \ref{sec:agnmodes}, we delve deeper into the specific impact of the positive AGN feedback and the hot gas ejection mode on galaxy evolution.

Given the importance of the topics addressed, it is worth reiterating the fundamental differences between \texttt{FEGA25} and its predecessor, \texttt{FEGA24}. The first significant update in \texttt{FEGA25} involves a more detailed description of stellar halos surrounding central galaxies (\citealt{contini2024e}). In this framework, these stellar halos form through the reincorporation of stars originally belonging to the diffuse light or intracluster light (ICL) in a transition region between the ICL and the central galaxy. However, this implementation does not influence the analysis conducted in this work.

The second major update is particularly important and represents an unprecedented advancement in semianalytic modeling. While \texttt{FEGA24} isolated the positive AGN feedback using Equation~\ref{eqn:pAGN}, \texttt{FEGA25} introduces a more physically-motivated approach, described by Equation~\ref{eqn:newpAGN}. This new formulation links the positive AGN feedback to its negative counterpart, both controlled by a single free parameter, the BH accretion efficiency, $\kappa_{\rm{AGN}}$.

The third, and most novel, addition to our SAM is the hot gas ejection mode, where AGN activity ejects gas beyond the virial radius of a halo, governed by Equation~\ref{eqn:ejection}. This mode introduces an additional parameter, $V_{\rm{scale}}$, which controls the amount of ejected hot gas, with stronger effects in less massive halos. Together, these innovations make \texttt{FEGA25} a more versatile and realistic model for galaxy evolution.

%%%%%%%%%%%%%%%%%%%%%%%%%%%%%%%%%%%%%%%%%%%%%%%%%%%%%%%%%%%%%%%%%%%%%%%%%%%%%%%%%%%%%%%%%%%%%%%%%%%%%%%%%%%%%%%%%%%%%%
\subsection{Evolution of the Stellar Mass Function}
\label{sec:smf}

\begin{figure*}[t!]
\begin{center}
\begin{tabular}{cc}
\includegraphics[width=0.35\textwidth]{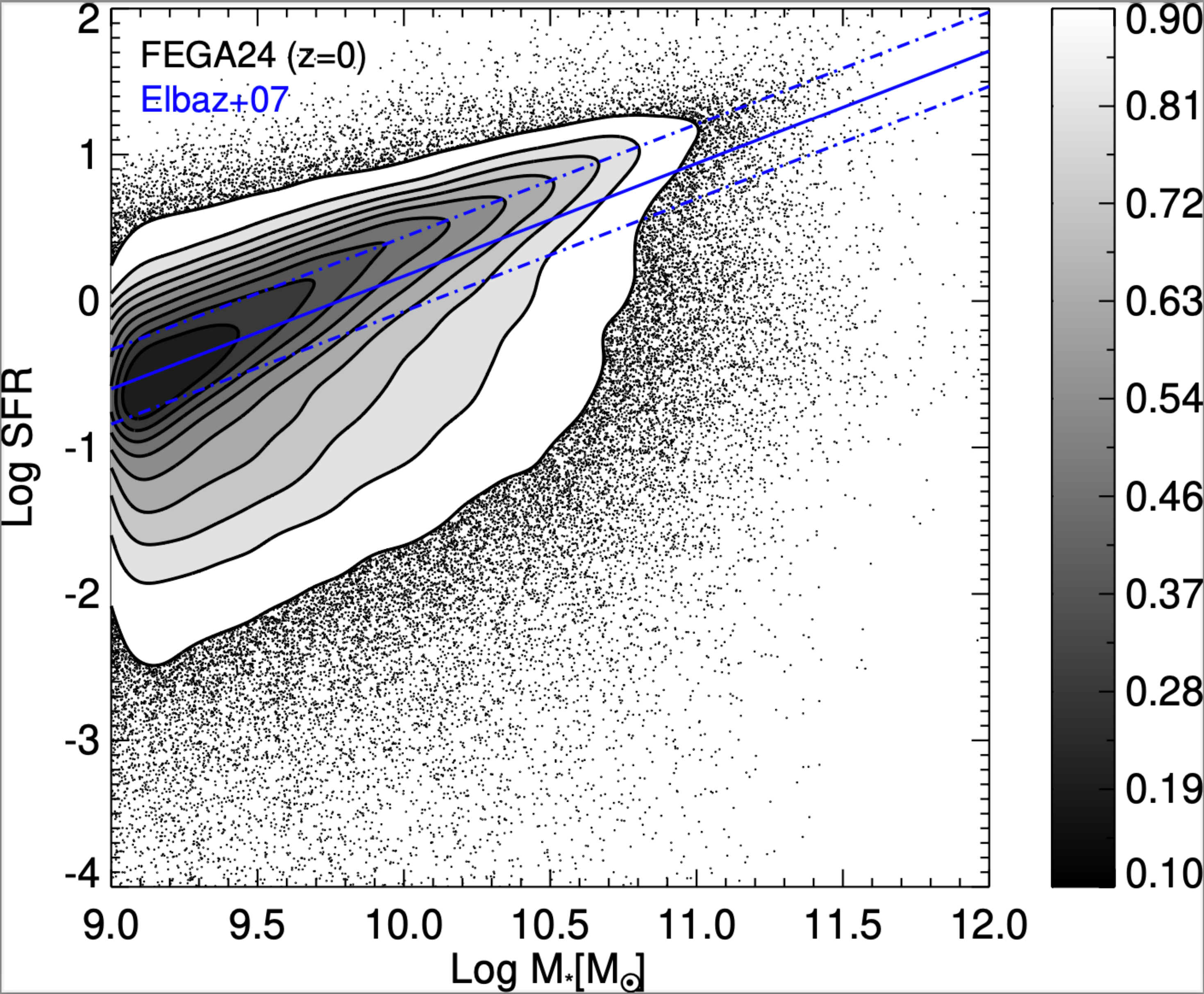} &
\includegraphics[width=0.35\textwidth]{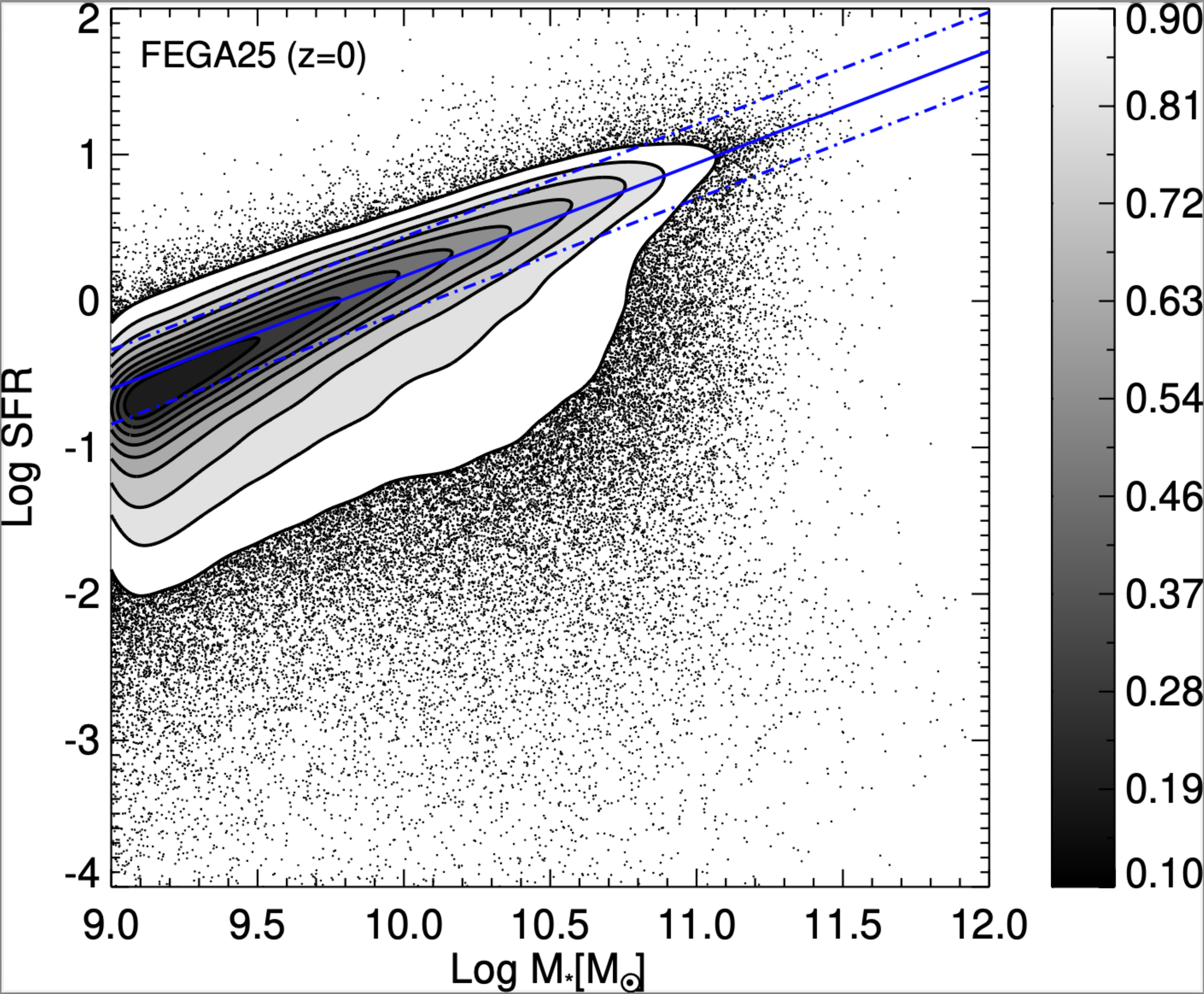} \\
\includegraphics[width=0.35\textwidth]{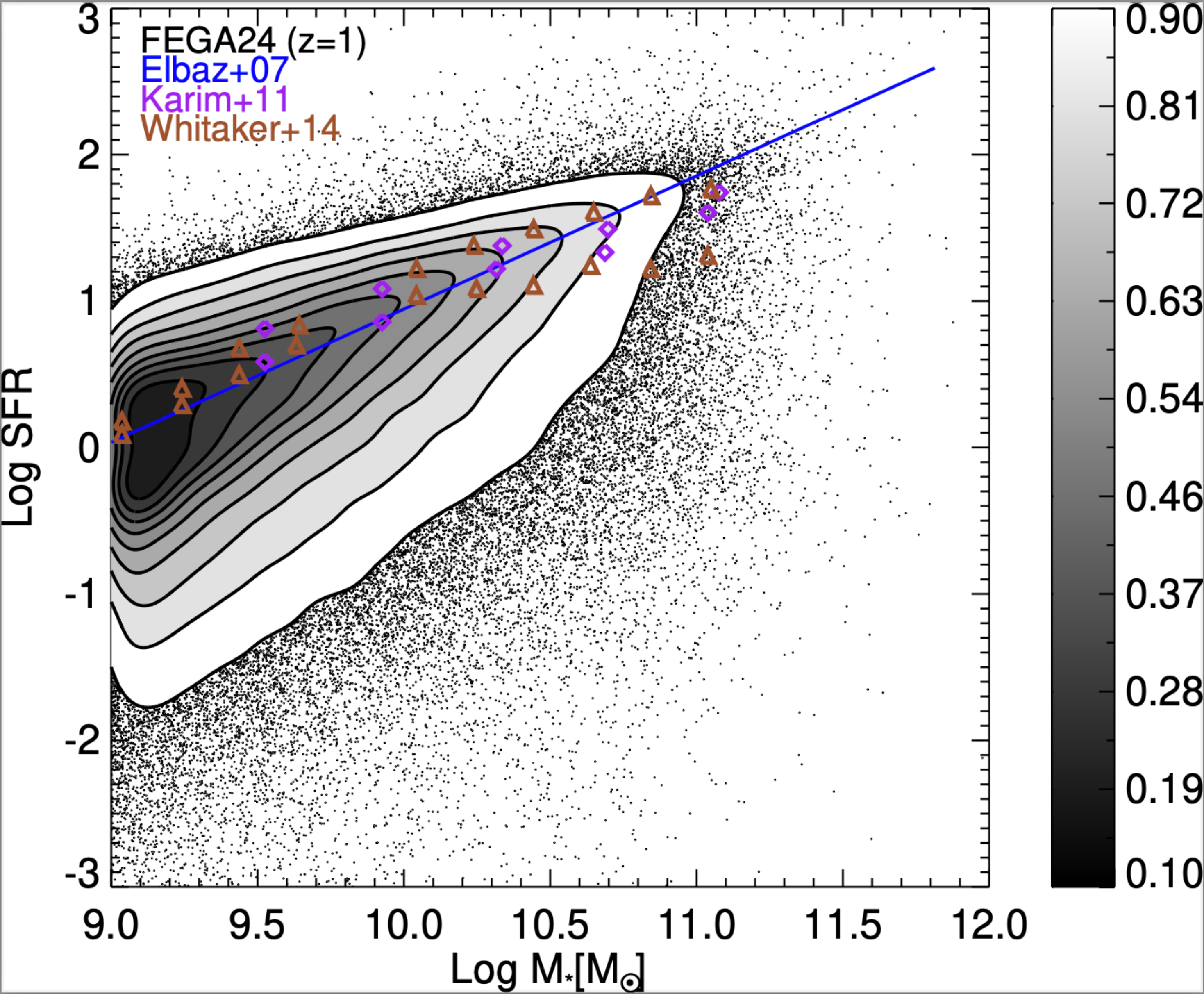} &
\includegraphics[width=0.35\textwidth]{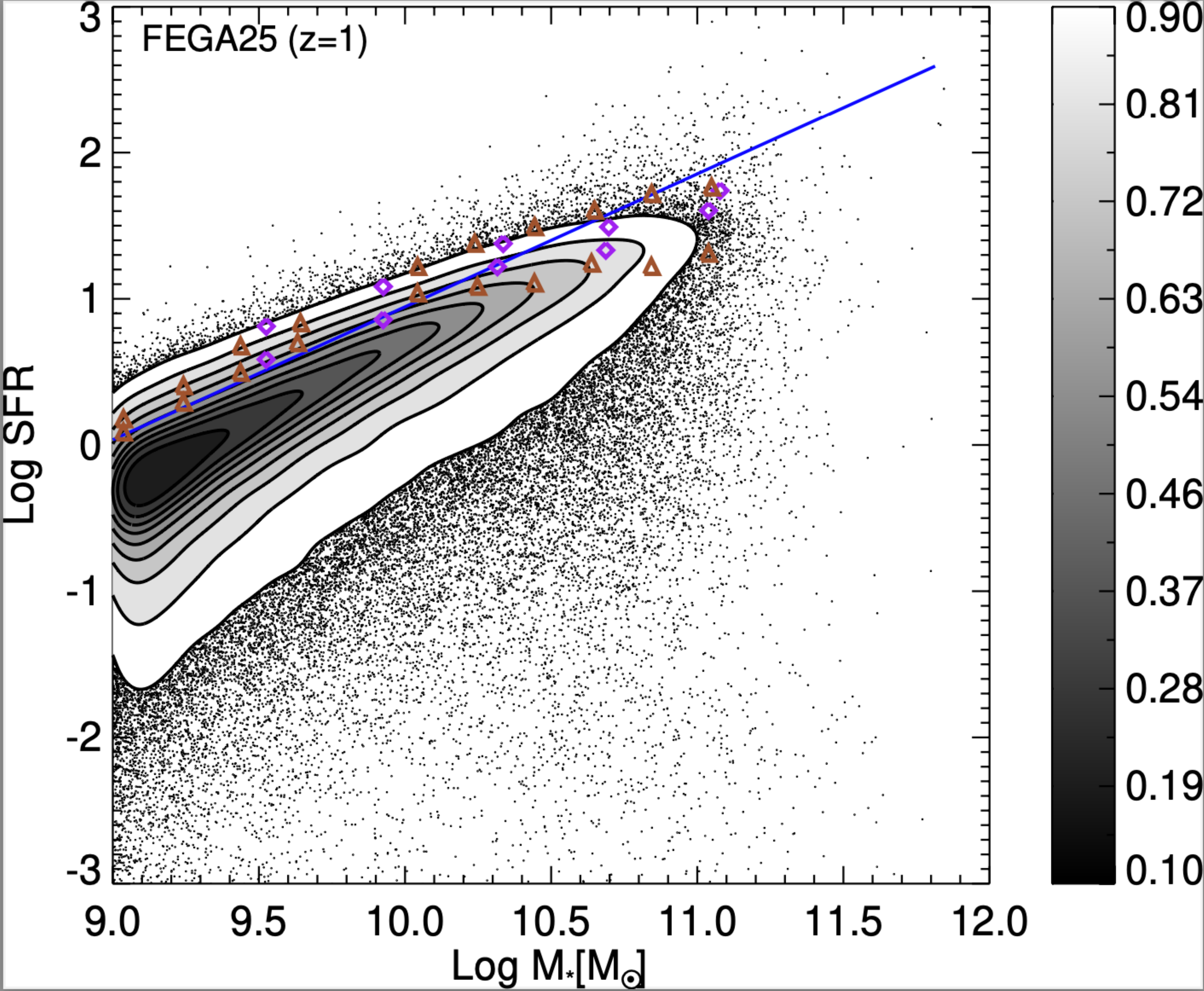} \\
\includegraphics[width=0.35\textwidth]{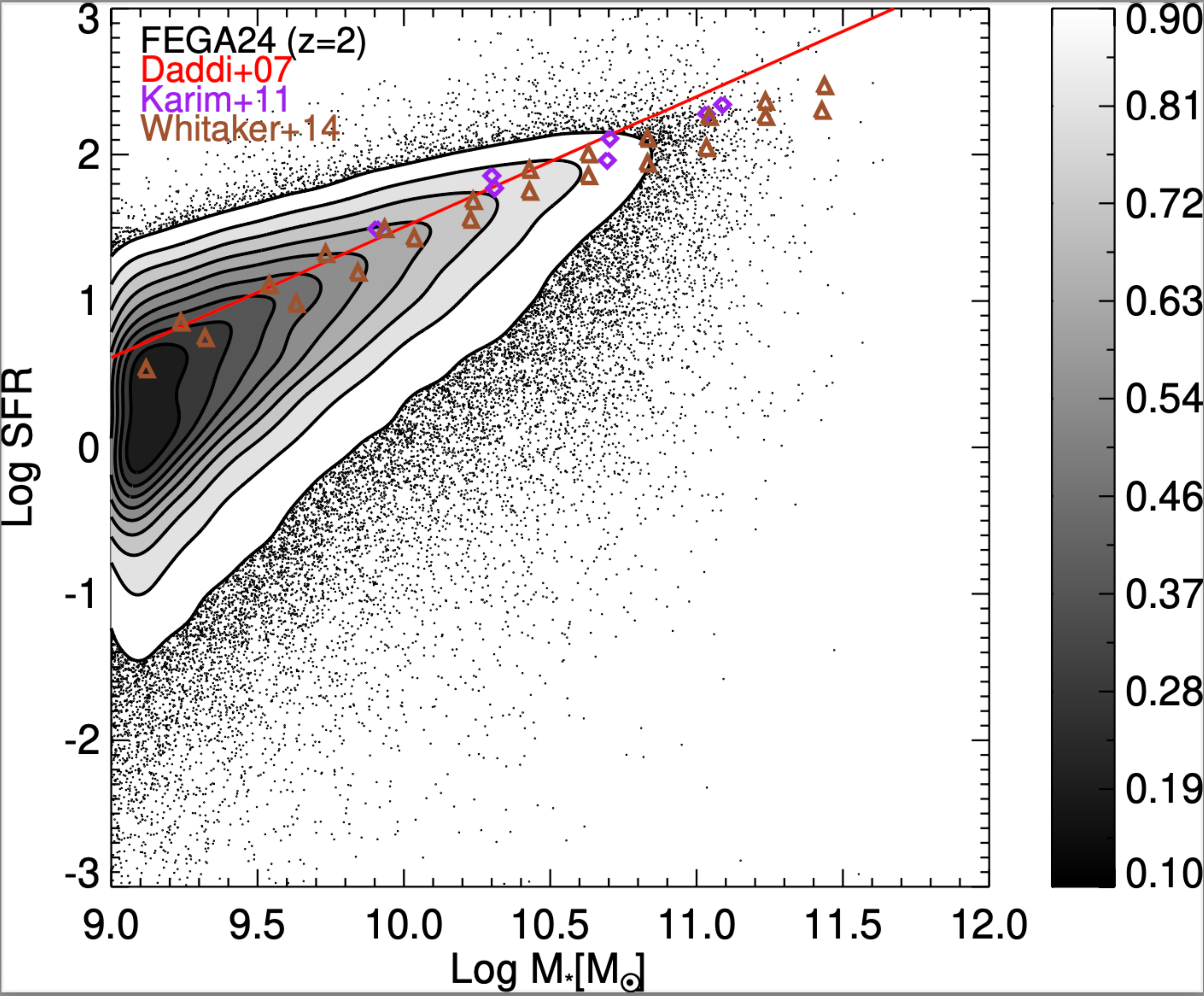} &
\includegraphics[width=0.35\textwidth]{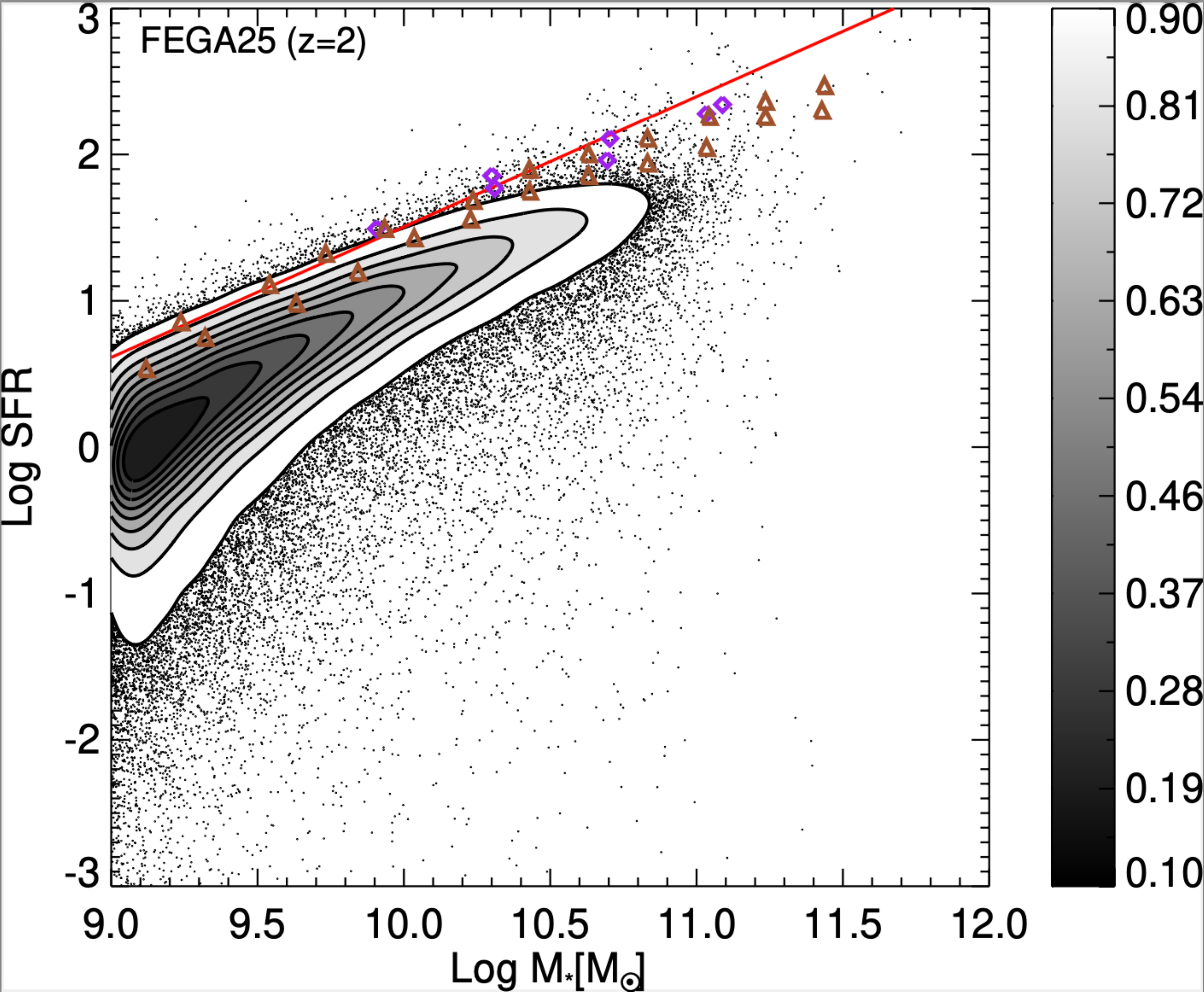} \\
\includegraphics[width=0.35\textwidth]{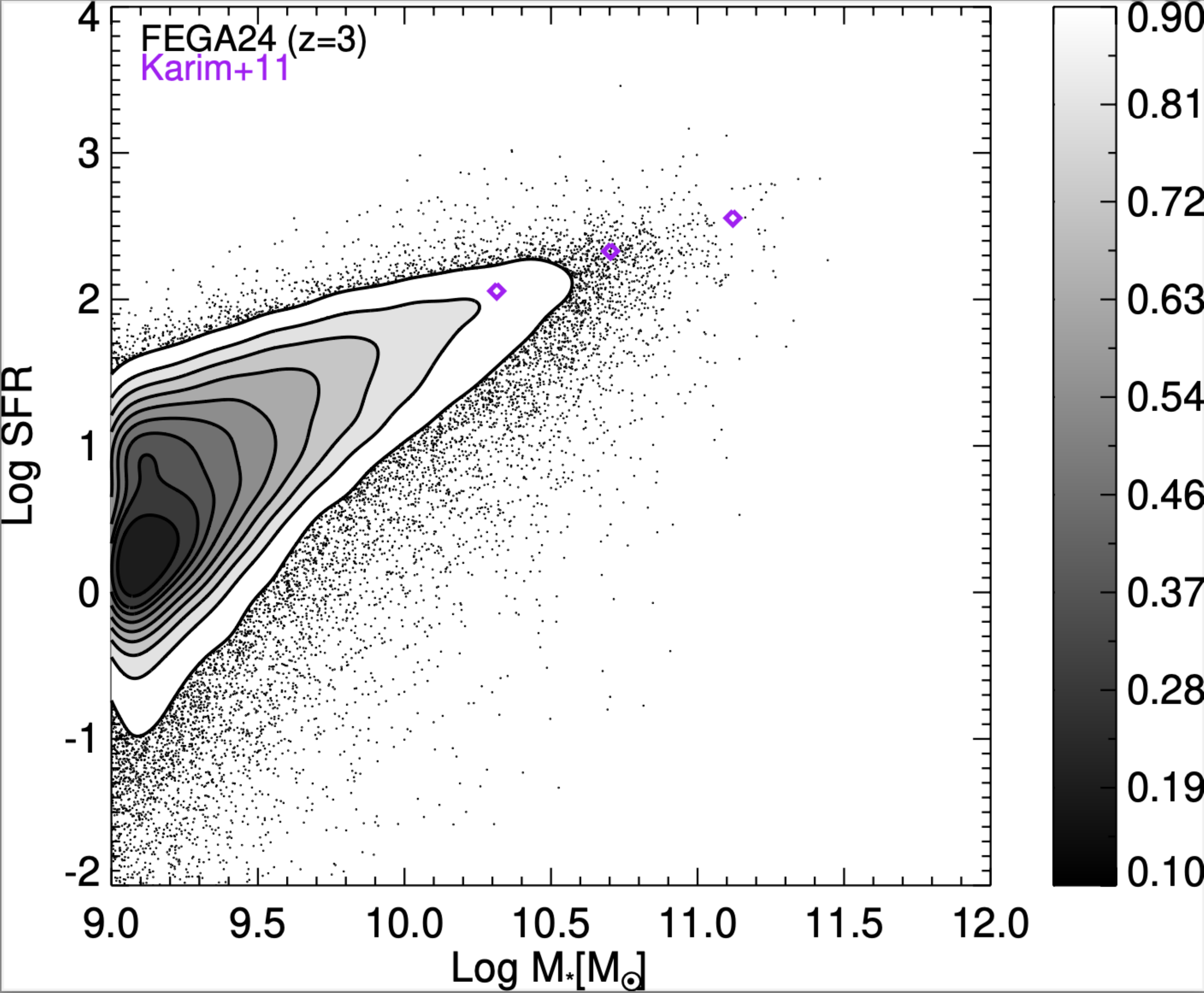} &
\includegraphics[width=0.35\textwidth]{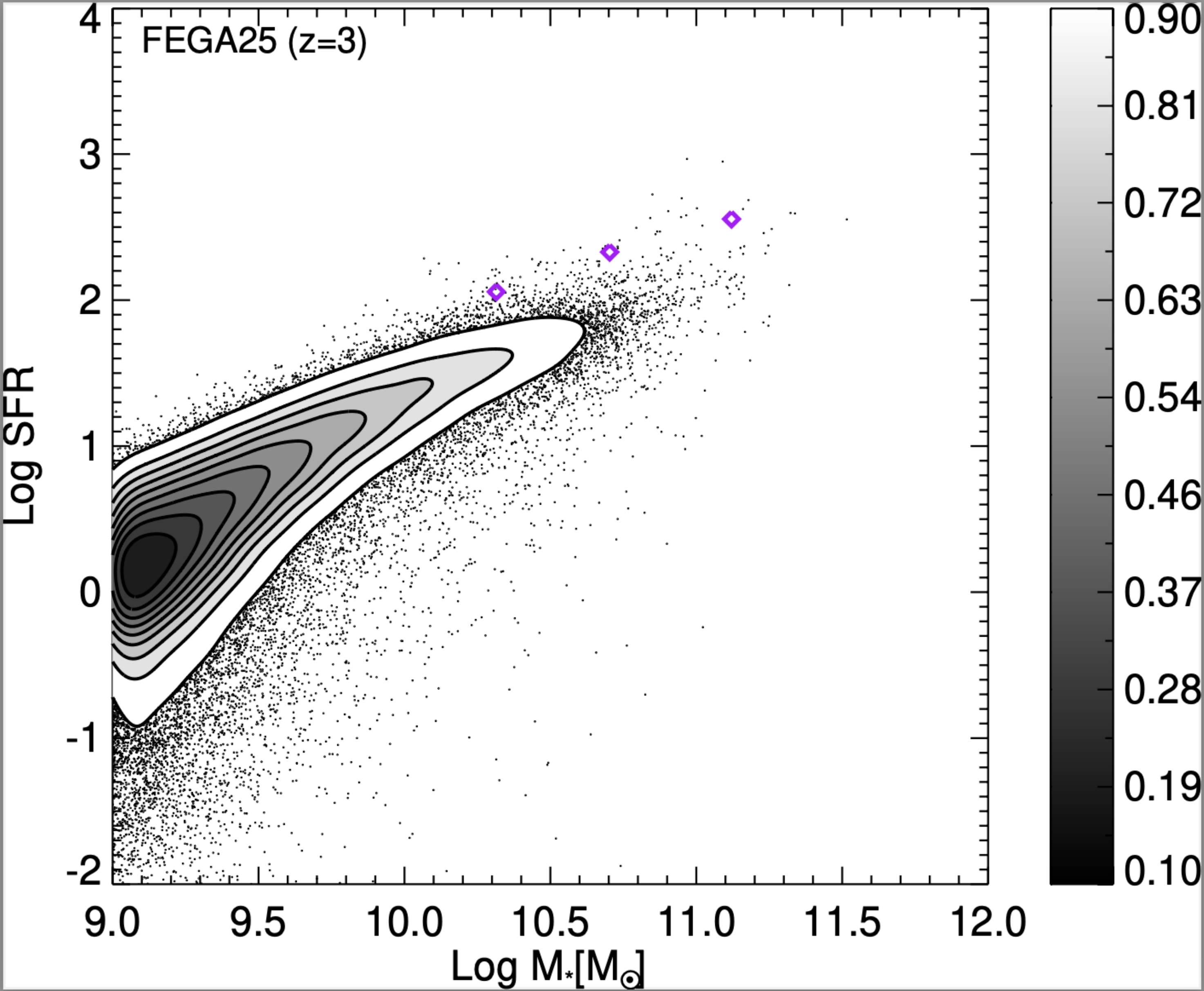} \\
\end{tabular}
\caption{The star formation rate-stellar mass relation at $z=0,1,2,3$ is shown, as predicted by \texttt{FEGA24} (left panels) and \texttt{FEGA25} (right panels). Both models are compared with the observed main sequence data from \citealt{elbaz2007} (represented by blue lines), \citealt{daddi2007} (blue lines), \citealt{karim2011} (purple diamonds) and \citealt{whitaker2014} (brown tringles). The contours represent different percentages of the data, as indicated by the colors in the accompanying bars. The last contour encloses 90\% of the full sample, while the remaining 10\% are depicted by individual black dots. Both models are able to catch the overall evolution of this relation, although (apparently, see text for more details) \texttt{FEGA24} seems to be more in line with observations.}
\label{fig:sfrmass}
\end{center}
\end{figure*}

We proceed with the analysis by examining Figure \ref{fig:SMFevol}, which depicts the evolution of the SMF from $z=3$ to the present day. The predictions from \texttt{FEGA25} (blue squares) are compared to those of \texttt{FEGA24} (green stars), the results from \texttt{L-Galaxies SAM} (\citealt{henriques2020}, brown dashed lines), and various observational datasets, as indicated in the legend. As in C24, we remind the reader that the evolution of the SMF was used to calibrate the model. Therefore, these predictions should not be considered fully independent.

The figure reveals that both models effectively capture the overall SMF evolution and are consistent with the observational data used during calibration. However, minor differences emerge at the high-mass end across all redshifts shown. Specifically, \texttt{FEGA25} predicts slightly lower number densities for massive galaxies compared to \texttt{FEGA24}.

The minor differences between the two models can be attributed to several factors. Although the BH accretion efficiency, $\kappa_{\rm{AGN}}$, is 2.5 times lower in \texttt{FEGA25}—which would suggest an opposite trend—the differing implementations of positive AGN feedback in the models, and particularly the new prescription for hot gas ejection in \texttt{FEGA25}, play a crucial role. Due to the ejection of some hot gas driven by AGN feedback, the amount of hot gas available for cooling in large groups and clusters, where AGN feedback is active, may be somewhat reduced. Consequently, $\kappa_{\rm{AGN}}$ could also decrease. The net effect is a lower overall efficiency, as the reduced availability of hot gas is sufficient to suppress cooling flows, which are also expected to be, on average, slightly diminished.

Furthermore, while the positive feedback mode contributes to the stellar mass of massive galaxies, the reduced availability of gas for cooling leads to less gas being available for star formation overall. The interplay of these mechanisms explains the evolution of the SMF observed in \texttt{FEGA25}, highlighting the complex and interconnected nature of the feedback processes.

That said, one caveat must be noted. As discussed in C24, the high-mass end of the SMF at high redshift is somewhat underpredicted due to the limited box size, which may not be large enough to capture a representative sample of massive galaxies at those redshifts. Additionally, it is important to highlight that our data reflect the raw number densities produced by the models, rather than fitted values.\

\begin{figure}[t!]
\centering
\includegraphics[width=0.45\textwidth]{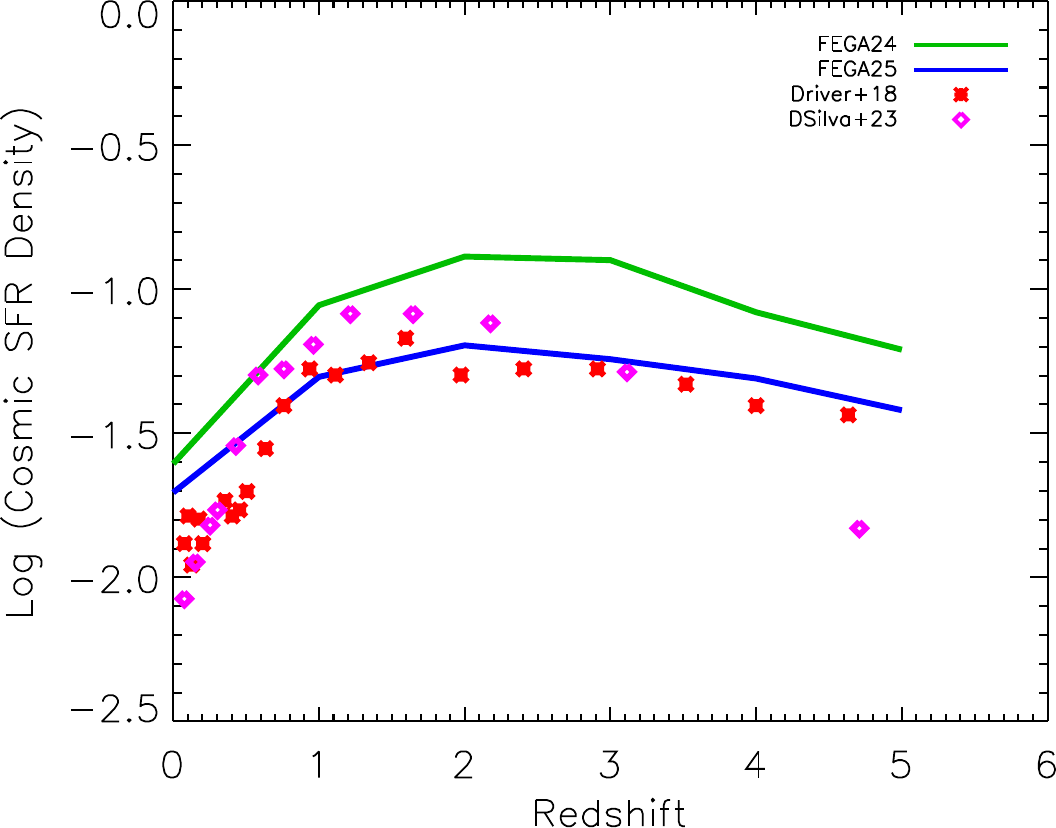}
\caption{Cosmic SFR density as a function of redshift as predicted by \texttt{FEGA24} (green line) and \texttt{FEGA25} (blue line), compared with the observed data from \cite{driver2018}, (red crosses) and
from \cite{dsilva2023} (magenta diamonds). \texttt{FEGA25} is more capable to follow the observed data than \texttt{FEGA24} at redshift $z>2$. This plot accounts for all galaxies in YS200 having $\log M_{*}\geq 8.0$.}
\label{fig:csfrd}
\end{figure}

\begin{figure*}[t!]
\begin{center}
\begin{tabular}{cc}
\includegraphics[width=0.47\textwidth]{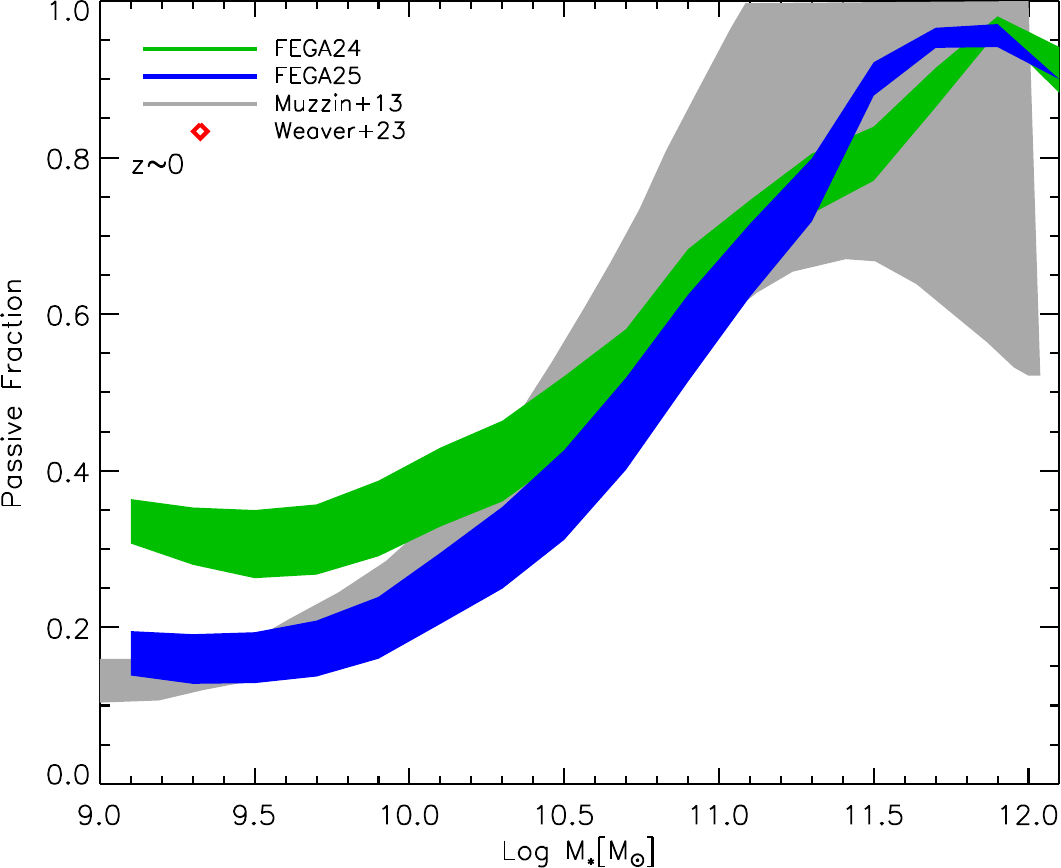} &
\includegraphics[width=0.47\textwidth]{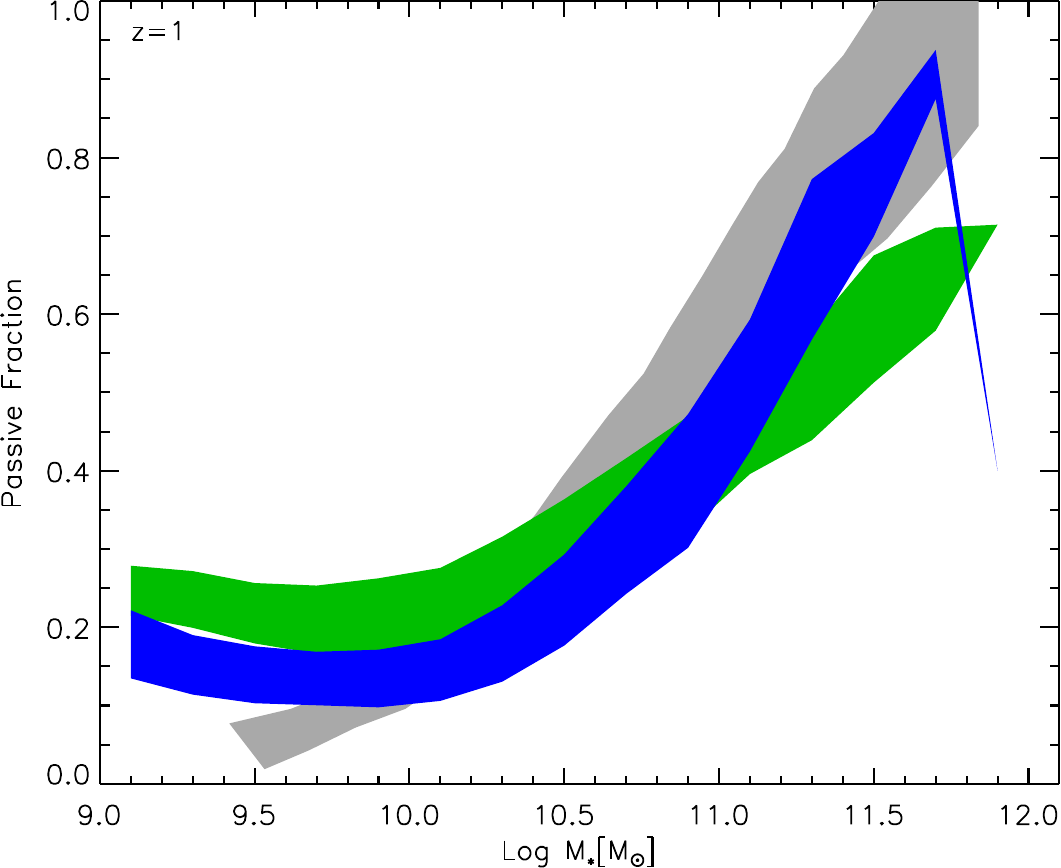} \\
\includegraphics[width=0.47\textwidth]{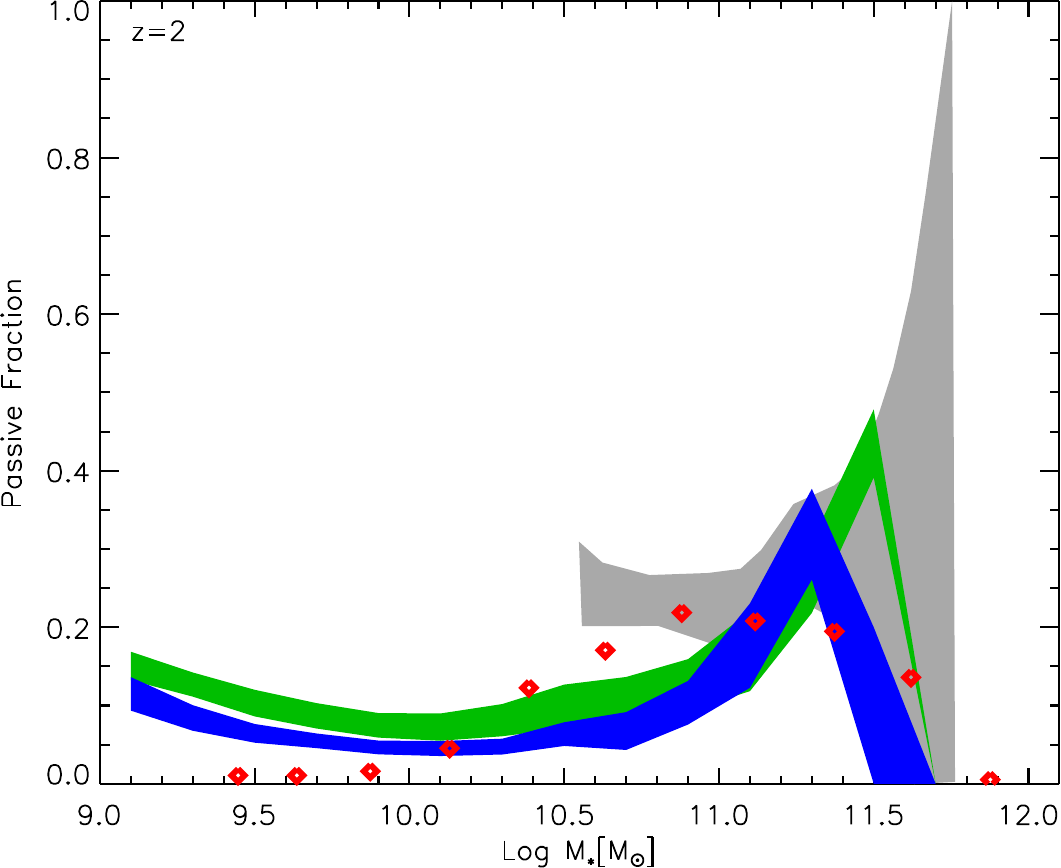} &
\includegraphics[width=0.47\textwidth]{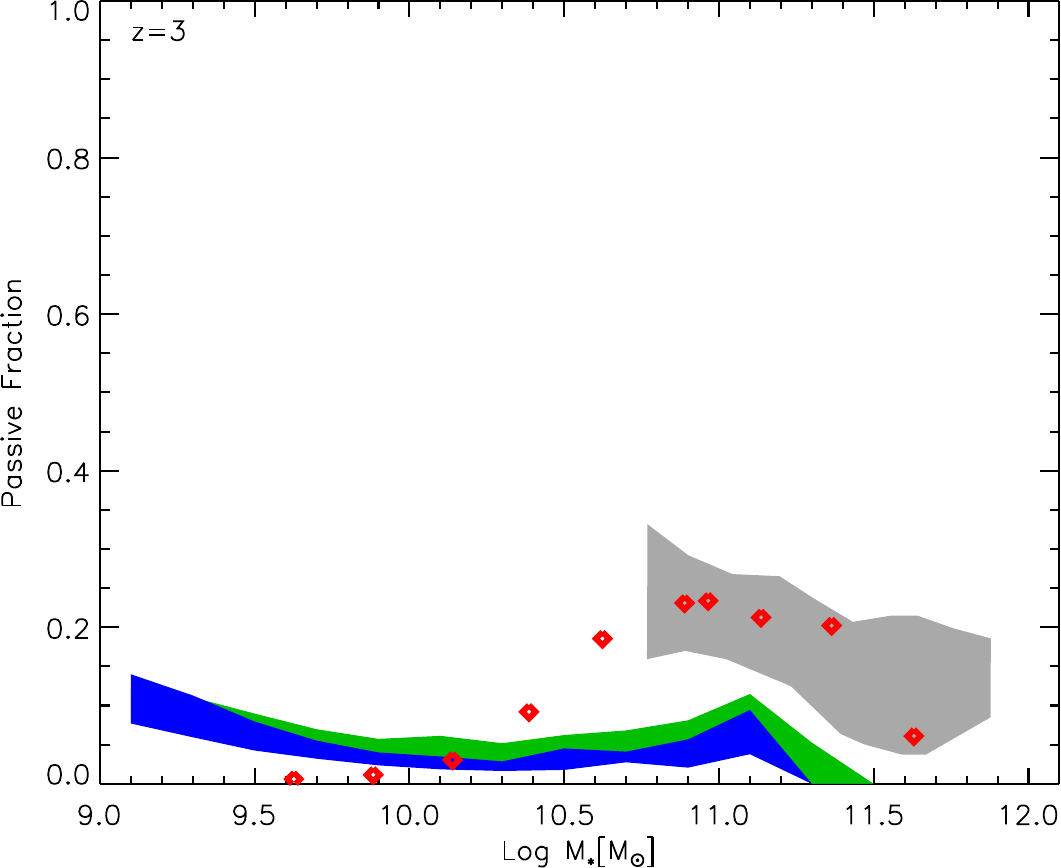}
\end{tabular}
\caption{The fraction of passive galaxies as a function of stellar mass, at $z=0.2,1,2,3$ as predicted by \texttt{FEGA24} (represented by the green regions) and \texttt{FEGA25} (represented by blue regions), compared withn the observational data from \citealt{muzzin2013} {green regions} and from \citealt{weaver2023} (red diamonds). Both model can describe the evolution with redshift of the passive fraction of galaxies up to $z=2$, with \texttt{FEGA25} being more accurate, but they both underestimate the observed fractions at $z=3$.}
\label{fig:redfraction}
\end{center}
\end{figure*}

%%%%%%%%%%%%%%%%%%%%%%%%%%%%%%%%%%%%%%%%%%%%%%%%%%%%%%%%%%%%%%%%%%%%%%%%%%%%%%%%%%%%%%%%%%%%%%%%%%%%%%%%%%%%%%%%%%%%%%
\subsection{FEGA24 vs FEGA25}
\label{sec:models}

% \begin{figure*}[t!]
% \begin{center}
% \begin{tabular}{cc}
% \includegraphics[width=0.47\textwidth]{figures/passive_fraction_z0.pdf} &
% \includegraphics[width=0.47\textwidth]{figures/passive_fraction_z1.pdf} \\
% \includegraphics[width=0.47\textwidth]{figures/passive_fraction_z2.pdf} &
% \includegraphics[width=0.47\textwidth]{figures/passive_fraction_z3.pdf}
% \end{tabular}
% \caption{{\bf The fraction of passive galaxies as a function of stellar mass, at $z=0.2,1,2,3$} as predicted by \texttt{FEGA24} (represented by the green regions) and \texttt{FEGA25} (represented by blue regions), compared withn the observational data from \citealt{muzzin2013} {green regions} and from \citealt{weaver2023} (red diamonds). Both model can describe the evolution with redshift of the passive fraction of galaxies up to $z=2$, with \texttt{FEGA25} being more accurate, but they both underestimate the observed fractions at $z=3$.}
% \label{fig:redfraction}
% \end{center}
% \end{figure*}

Before delving into the details of the analysis—namely, the genuine predictions of \texttt{FEGA25} and comparisons with its predecessor and observational data—it is important to highlight a key point. In the analysis presented in C24, which examined various galaxy properties and scaling relations, \texttt{FEGA24} showed a good level of reproduction. However, \texttt{FEGA25} not only achieves similar results but also introduces notable improvements on some of them, which will be the focus of the following discussion.

Figure~\ref{fig:sfrmass} presents the star formation rate (SFR)–stellar mass relation at $z=0,1,2,3$, as predicted by \texttt{FEGA24} (left panels) and \texttt{FEGA25} (right panels), both compared to several observations as indicated by the legend. Overall, both models successfully reproduce the main sequence at $z=0$ and also provide a reasonable depiction of the broad red cloud. However, there is some difference between them. Specifically, while the main sequence predicted by \texttt{FEGA24} is slightly biased high relative to the observed one—albeit still within the observational scatter—the main sequence predicted by \texttt{FEGA25} aligns perfectly with the observed average (the contours represent different percentages of the data, as indicated by the colors in the accompanying bars, and the last contour encloses 90\% of the full sample, while the remaining 10\% are represented by individual black dots). This improvement is a direct outcome of the new AGN feedback implementation in \texttt{FEGA25}, which more effectively regulates the availability of hot gas compared to the mechanism employed in \texttt{FEGA24}.

At higher redshifts, the main sequence predicted by \texttt{FEGA24} appears to be more consistent with observational data than that produced by \texttt{FEGA25}, which shows a closer alignment with other SAMs, such as \texttt{L-Galaxies} (\citealt{henriques2020}, their Figure 7). However, it is worth noting that the main sequence from \texttt{FEGA24} is generally broader compared to that of \texttt{FEGA25}. As we will discuss in more detail later, this broader distribution is a direct consequence of the stronger implementation of positive AGN feedback in \texttt{FEGA24}.

The overall evolution of the SFR is best represented by the cosmic SFR density as a function of redshift, shown in Figure \ref{fig:csfrd}. This provides a crucial benchmark for evaluating how well our models capture the buildup of stellar mass over cosmic time. In this figure, we compare the predictions of our models—depicted by the green line for \texttt{FEGA24} and the blue line for \texttt{FEGA25}—against observational data from \citealt{driver2018} (red crosses) and \citealt{dsilva2023} (magenta diamonds).

A key result from this comparison is that \texttt{FEGA24} successfully reproduces the cosmic SFR density only in the last $\sim$7 Gyr, particularly when compared to the dataset from \cite{dsilva2023}. However, at higher redshifts, its predictions deviate from the observational constraints, indicating that the original model struggles to fully capture the early phases of star formation history. In contrast, \texttt{FEGA25} shows a much-improved agreement with the observed cosmic SFR density, particularly with the dataset from \cite{driver2018}. Notably, \texttt{FEGA25} maintains a close match up to $z=5$, suggesting that the refinements in the new model allow for a more accurate representation of early galaxy formation and evolution.

Another significant improvement over the previous model is illustrated in Figure~\ref{fig:redfraction}, which shows the fraction of passive galaxies as a function of their stellar mass, and at different redshift (different panels). The predictions from \texttt{FEGA25} are shown in blue, while those from \texttt{FEGA24} are shown in green. Both are compared with observed fractions derived from \citealt{muzzin2013} (grey regions) and from \citealt{weaver2023} (red diamonds). Following C24, passive galaxies are defined consistently with the above-mentioned observations, based on different thresholds in the logarithm of the specific star formation rate (SSFR), set to be $0.3\cdot t^{-1}_{\rm{Hubble}}$ (around $10^{-11} {\rm yr}^{-1}$ at $z=0$). The green and blue regions represent the range encompassing the lowest and highest ($+0.5$ dex) thresholds in $\log \text{SSFR}$.

Figure~\ref{fig:redfraction} shows both model can describe the evolution with redshift of the passive fraction of galaxies up to $z=2$, with \texttt{FEGA25} being more accurate, but they both underestimate the observed fractions at $z=3$. \texttt{FEGA25} reproduces the passive fraction of galaxies in greater detail than \texttt{FEGA24} up to $z=2$. This is evident not only in the overall trend of an increasing passive fraction with stellar mass, but particularly in its ability to get closer to the observed fractions at lower stellar masses. Interestingly, the gap between the models in the passive fraction (at $z=0,1,2$) at low stellar masses—although counterintuitive—is a direct consequence of the introduction of the hot gas ejection mode. Without this mechanism, \texttt{FEGA25} would behave similarly to \texttt{FEGA24}, despite the differences in their positive feedback implementations. In low-mass halos, the ejection of gas appears to result in a larger fraction of dwarf galaxies remaining active \footnote{This feature is caused by the introduction of the hot gas ejection mode. In fact, when \texttt{FEGA25} is run with the new implementation of positive AGN feedback alone, the model still behaves similarly to the original version.}, likely due to the increased availability of cold gas (this point is under investigation in Contini et al. in prep).

\begin{figure*}[t!]
\centering
\includegraphics[width=0.9\textwidth]{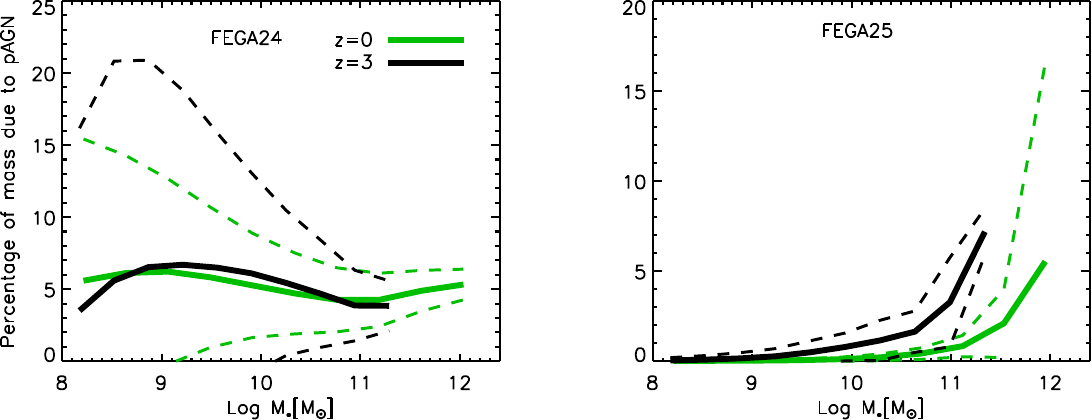}
\caption{As done in C24, we present the percentage of stellar mass contributed by the positive AGN feedback mode as a function of stellar mass (for central galaxies only). The predictions of \texttt{FEGA24} (left panel) and \texttt{FEGA25} (right panel) are shown at two redshifts: $z=0$ (green lines) and $z=3$ (black lines). The plots account for all stellar mass attributed to positive AGN feedback by considering all branches of the merger tree for each galaxy. There are notable differences between the positive AGN feedback implemented in C24 (left panel) and the new implementation in \texttt{FEGA25} (right panel). First, the new positive AGN feedback in \texttt{FEGA25} naturally and distinctly captures the expected redshift dependence, with higher efficiency at higher redshifts. Secondly, while in C24 there was no clear trend with stellar mass, only a larger scatter at lower mass scales (left panel), \texttt{FEGA25} shows an increase in the efficiency of positive AGN feedback with stellar mass. This is particularly evident where the negative mode is also efficient, though not strong enough to heat all the cooling gas.}
\label{fig:pAGNcontr}
\end{figure*}

\begin{figure*}[t!]
\centering
\includegraphics[width=0.9\textwidth]{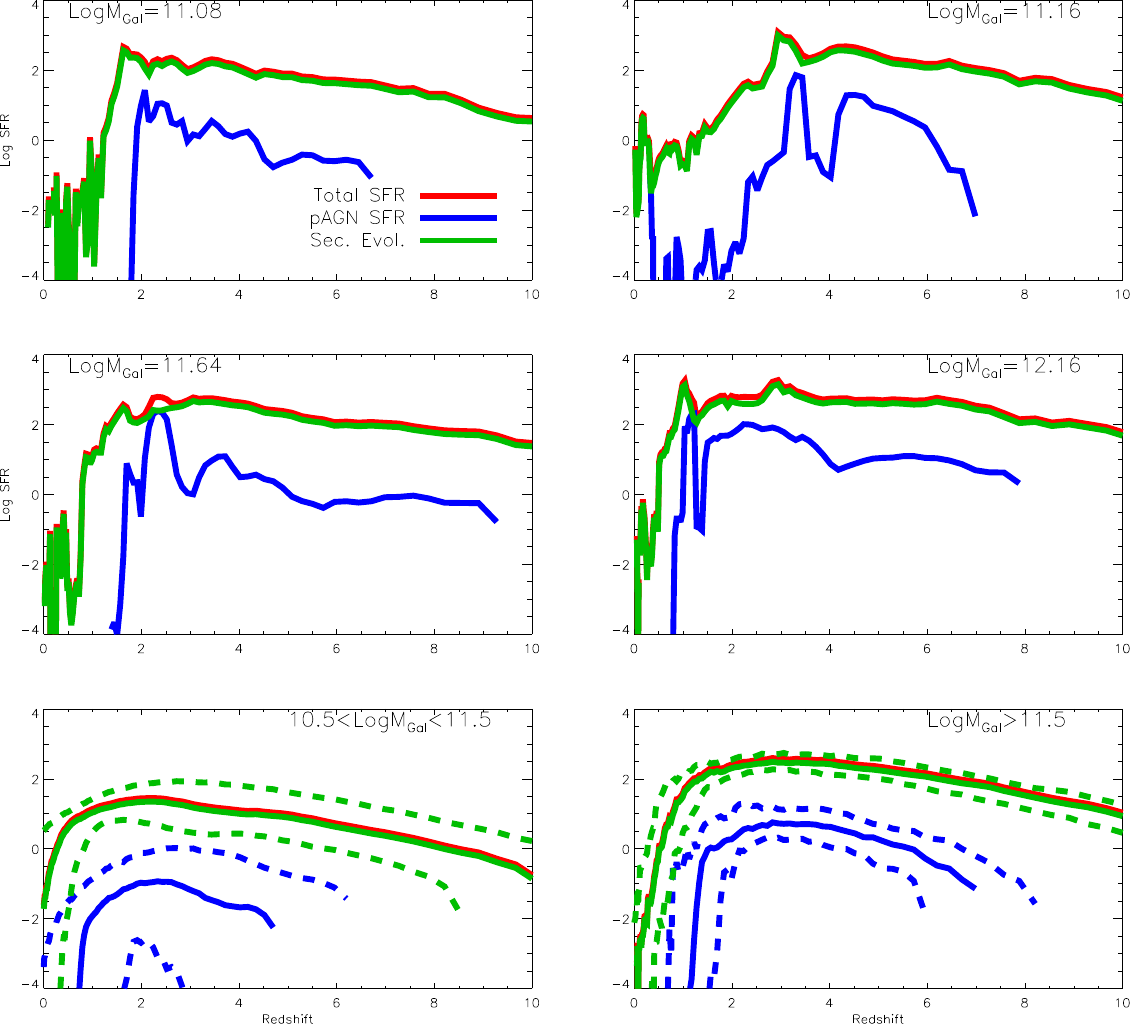}
\caption{The star formation history of four sample galaxies (top 4 panels), ordered by increasing stellar mass from top to bottom, is shown from redshift $z \sim 10$ to the present time. The green line represents the secular evolution, the blue line indicates the SFR due solely to positive AGN feedback, while the red one represent the total SFR (shifted 0.1 dex upword for easier viewing). Several interesting features can be observed in these plots. From top to bottom, it is clear that the contribution of positive AGN feedback to the total SFR increases, as expected from Figure~\ref{fig:pAGNcontr}. This trend is evident not only in terms of the mass provided but also in the duration of activity of the positive feedback mode. Specifically, in lower-mass galaxies, the contribution is shorter and characterized by episodic bursts, whereas in higher-mass galaxies, the SFR history driven by positive AGN feedback more closely follows the total SFR, being rather continuous than broken or irregular, being longer and more sustained. In all cases, the positive feedback mode is efficient at high redshift and begins to fade as the negative feedback mode takes over the cooling process, indicating that the negative mode reaches its peak power. The two bottom panels show, instead, the distributions (median, 16th and 84th percentiles) of the SFR histories for wide samples of galaxies in the stellar mass range $10.5< \log M_* <11.5$ (bottom left panel), and $\log M_* >11.5$ (bottom right panel).}
\label{fig:SFRhist}
\end{figure*}

\begin{figure*}[t!]
\centering
\includegraphics[width=0.8\textwidth]{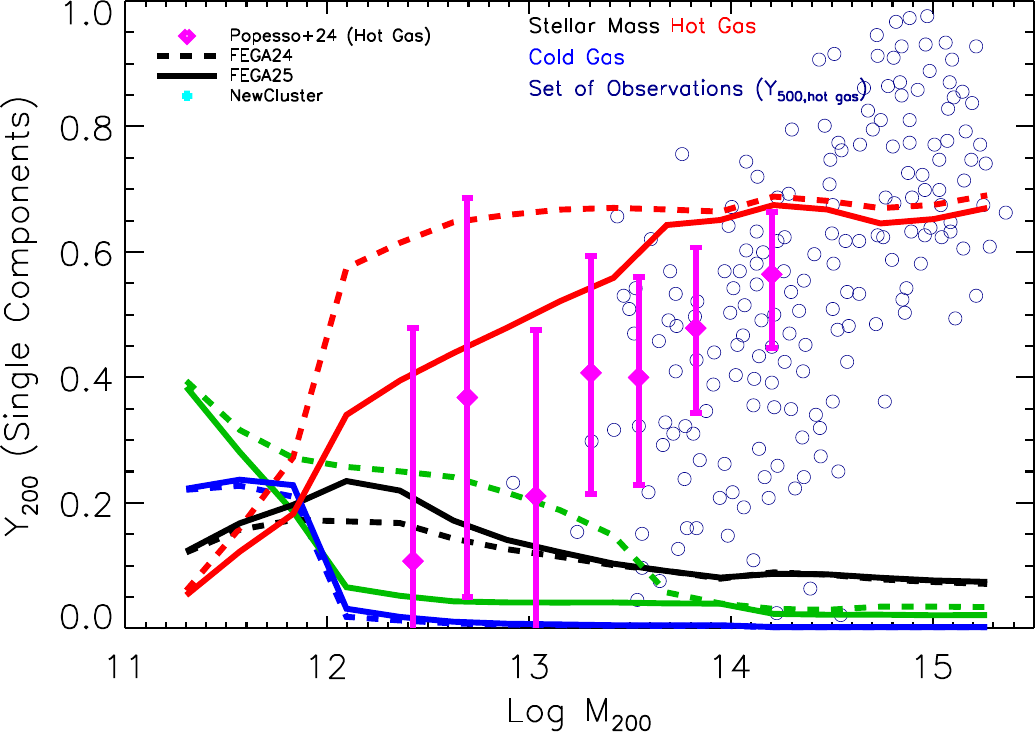}
\caption{The fraction of mass for different components (stars, cold gas, hot gas, and ejected gas) within the virial radius, normalized by the universal baryon fraction, is shown as a function of halo mass. Different colors represent the various components, and the dashed and solid lines distinguish the predictions from \texttt{FEGA24} and \texttt{FEGA25}, respectively. Our models are compared with the observed hot gas fraction from \citealt{popesso2024} (magenta diamonds with error bars) and a set of observed hot gas fractions from various authors (\citealt{vikhlinin2006,arnaud2007,sun2009,rasmussen2009,pratt2009,mahdavi2013,gonzalez2013,lovisari2015,eckert2016,pearson2017,eckert2019,mulroy2019,ragagnin2022}, dark blue circles).
Given the importance of sampling halos down to $\log M_{\rm{halo}} = 11$, for this plot we also make use of the catalogs obtained by running the models on the merger trees extracted from YS50HR.
It is important to note that, despite the large scatter in the observed data, \texttt{FEGA24} fails to predict a reasonable hot gas fraction for halos smaller than large groups ($\log M_{\rm{halo}} < 14$). In contrast, \texttt{FEGA25}, with its new hot gas ejection mode in AGN feedback, performs significantly better. This result underscores the importance of having a mechanism, such as the hot gas ejection mode, that can expel gas from the halo without significantly altering the cold gas and stellar mass content.}
\label{fig:compF_halomass}
\end{figure*}

%%%%%%%%%%%%%%%%%%%%%%%%%%%%%%%%%%%%%%%%%%%%%%%%%%%%%%%%%%%%%%%%%%%%%%%%%%%%%%%%%%%%%%%%%%%%%%%%%%%%%%%%%%%%%%%%%%%%%%
\subsection{Positive AGN feedback and Hot Gas Ejection Modes}
\label{sec:agnmodes}

In Figure~\ref{fig:pAGNcontr}, we highlight the differences in the positive AGN feedback mechanisms implemented in the two models. The figure shows, for central galaxies only, the percentage of stellar mass attributed to positive AGN feedback as a function of stellar mass, as predicted by \texttt{FEGA24} (left panel) and \texttt{FEGA25} (right panel) at $z=3$ (black lines) and $z=0$ (green lines). As in C24, this calculation includes all stellar mass generated by positive AGN feedback, accounting for all branches in the merger tree of each galaxy.

The most evident difference between the two models lies in their trends with stellar mass. In \texttt{FEGA24}, the relation is flat, with a larger scatter for lower-mass galaxies and an average contribution of approximately 5\%. In contrast, \texttt{FEGA25} exhibits an increasing percentage and scatter with stellar mass, starting with virtually no contribution for dwarf and intermediate-mass galaxies and rising to about 5\% for massive galaxies at $z=0$. At $z=3$, the percentages in \texttt{FEGA24} remain consistent, albeit with larger scatter, while \texttt{FEGA25} shows an increasing contribution toward lower-mass galaxies (with respect to $z=0$). This highlights a key distinction between the models: in \texttt{FEGA24}, redshift dependence is mildly reflected only through the scatter, whereas in \texttt{FEGA25}, it is a clear and systematic trend.

This result is particularly significant because it aligns well with recent theoretical predictions (e.g., \citealt{silk2024}), which suggest that positive AGN feedback is stronger at high redshifts—when negative feedback mechanisms are less active—and diminishes at lower redshifts, as negative feedback becomes dominant. Nevertheless, given the limited observational evidence for positive feedback (see references above), especially at high redshifts, it is important to emphasize that neither implementation can be definitively ruled out. Both positive and negative feedback modes may coexist. The advantage of the \texttt{FEGA25} prescription is its ability to accurately describe the redshift dependence of positive AGN feedback.

To gain an understanding of the contribution of positive AGN feedback to star formation in \texttt{FEGA25}, we present the SFR histories of four galaxies with increasing stellar mass, from the four top panels, in Figure~\ref{fig:SFRhist}. The green line represents the secular evolution, the blue line indicates the SFR due solely to positive AGN feedback, while the red one represent the total SFR (shifted 0.1 dex upword for easier viewing). As expected from the results in the previous figure, the overall contribution of the positive feedback mode to star formation increases with the galaxy’s stellar mass, both in terms of the mass it provides and the duration for which the positive feedback is active.

The SFR history driven by positive AGN feedback closely tracks the total SFR, displaying a rather smooth and continuous behavior rather than a broken or irregular pattern. In contrast, the total SFR is primarily sustained by secular evolution. Across all the representative cases examined, the positive feedback mode is most efficient at high redshift and gradually fades as the negative feedback mode becomes dominant, suppressing cooling more effectively. The exact redshift at which this transition occurs varies across different systems, but in general, the efficiency of the positive mode drops significantly after $z=2$, with only occasional bursts seen at later times. Notably, in several cases, the SFR triggered by positive feedback can reach—and in some instances even exceed—100 $M_{\odot}/\rm{yr}$. The bottom panels in Figure~\ref{fig:SFRhist}, instead, show the overall distributions of the SFR histories of galaxies, including the contribution of positive AGN feedback, in two different stellar mass ranges, $10.5< \log M_* <11.5$ (bottom left panel), and $\log M_* >11.5$ (bottom right panel).

Finally, Figure~\ref{fig:compF_halomass} serves as the key figure of this study, clearly illustrating the role of our hot gas ejection mode. The plot shows, for each baryonic component (denoted by different colors as described in the legend), the fraction of mass relative to the total mass within the virial radius, normalized by the universal baryon fraction, $\rm{Y}_{200}$, as a function of halo mass. Dashed lines represent the predictions of \texttt{FEGA24}, while solid lines correspond to those of \texttt{FEGA25}. The hot gas fraction predicted by our models is compared with recent results from \cite{popesso2024}, shown as magenta squares, and a set of observational data from different authors (see caption), represented by dark blue circles. It is important to note that, in the latter case, the fraction was computed within the virial region using an overdensity of $\Delta=500$ (with halo masses converted to $M_{200}$ in the plot), rather than $\Delta=200$ as in our models, so these values should be considered as lower limits compared to ours.

Figure~\ref{fig:compF_halomass} conveys a wealth of information, highlighting substantial differences between the models. Focusing first on the stellar mass (black lines), we observe that the peak of star formation occurs in Milky Way-like halos with $\log M_{\rm{halo}}=12.3$ for both models, although \texttt{FEGA25} predicts a slightly higher value. The cold gas fraction (blue lines) is around 20\% for $\log M_{\rm{halo}}<12.0$, then drops to zero as the stellar mass fraction reaches its peak. This trend applies to both models, with the only difference being that \texttt{FEGA25} predicts slightly higher cold gas fractions before the drop.

The key, and very important, distinction lies in the hot gas fraction (red lines), and by extension, in the ejected gas fraction (green lines). \texttt{FEGA24} (dashed line), which lacks a hot gas ejection mode in its AGN feedback, predicts a hot gas fraction of approximately 60\%-65\% for halos with $\log M_{\rm{halo}}>12.2$, from small groups to large clusters. In contrast, \texttt{FEGA25} (solid line), thanks to its hot gas ejection mode driven by AGN feedback, shows a considerably different hot gas fraction, more closely aligned with the observed data. The fraction of hot gas is noticeably reduced from low-mass to large groups, and while it matches the observations from \cite{popesso2024} at the edge of the scatter in several data points, it represents a clear improvement over \texttt{FEGA24}.

Since the cold gas and stellar components remain largely unchanged, the excess gas is directed into the ejecta (green lines). This behavior aligns with the desired outcome, as highlighted by other authors (see the discussion in \citealt{popesso2024}) and as discussed in Section~\ref{sec:intro}. The hot gas ejection mode in the AGN feedback has proven to be effective in reducing the hot gas within the virialized region.
As emphasized in Section~\ref{sec:intro} and reiterated here, this mechanism must effectively expel gas beyond the halo's virial radius without disrupting the internal processes that govern the cold gas and stellar components, ensuring the preservation of galaxy properties already well described by the model. That said, we observed minor differences in the amount of cold gas and stars in low-mass halos, which notably enhance the model’s ability to capture the fraction of passive dwarf galaxies.

It is clear that the new achievements in our SAM stem from the introduction of a more comprehensive AGN feedback model. The idea that AGN can expel gas beyond the halo’s virial radius is almost novel in SAMs (see SHARK2, \citealt{lagos2024}), though it has been quantified in numerical simulations (e.g., \citealt{pillepich2019,angelinelli2022,ayromlou2023,popesso2024}, among the most recent studies). \cite{popesso2024} provides an updated and detailed analysis of this phenomenon, comparing numerical simulation results with observational data (e.g., eROSITA eFEDS field data, \citealt{brunner2022}). As discussed in Section~\ref{sec:intro}, most simulations struggle to reproduce the observed fraction of hot gas within the virialized region over a broad range of halo masses, particularly at group scales. This shortfall underscores the need for an additional mechanism capable of pushing hot gas beyond the virial radius.

However, further investigation is needed for a particular reason. While \texttt{FEGA25} performs considerably better than its predecessor, its predictions still tend toward the upper limits of the observed range, suggesting that an even more efficient mechanism may be required. As described in Section~\ref{sec:ejection}, we can assume that the residual energy, after reheating the cooling gas in the negative feedback mode, is sufficient to expel some of the hot gas beyond the virial radius. While this is a plausible and valuable approach, it represents an extreme scenario that warrants a more in-depth investigation. We plan to explore this topic further in a forthcoming paper (Contini et al., in prep.), which will focus on the evolution of the baryon content in halos and its dependence on cosmic time.

%%%%%%%%%%%%%%%%%%%%%%%%%%%%%%%%%%%%%%%%%%%%%%%%%%%%%%%%%%%%%%%%%%%%%%%%%%%%%%%%%%%%%%%%%%%%%%%%%%%%%%%%%%%%%%%%%%%%%%%%
\section{Conclusions}
\label{sec:conclusion}

Building upon the semianalytic model introduced in \cite{contini2024d} and later updated in \cite{contini2024e}, we developed a new version named \texttt{FEGA25}, which introduces a comprehensive revision of AGN feedback, now split into three distinct modes: negative, positive, and hot gas ejection feedback. The first mode is the classical mechanism used in semianalytics and numerical simulations, which prevents hot gas from overcooling, thereby inhibiting star formation. The second is a more physically-motivated version of the mechanism presented in \cite{contini2024d}, accounting for star formation induced by AGN activity. The third mode represents the primary novelty of this work: a mechanism capable of expelling hot gas beyond the virial radius of the halo.

\texttt{FEGA25} successfully reproduces the galaxy properties and scaling relations explored in \cite{contini2024d}, while delivering significant improvements in the description of the main sequence of star-forming galaxies and the passive fraction as a function of stellar mass. The model suggests a positive feedback mode that is more active at high redshifts, gradually fading over time as the negative mode becomes dominant. On average, the positive mode is more effective in galaxies with increasing stellar mass. Thanks to the introduction of the third feedback mode—the hot gas ejection mode—the fraction of hot gas in halos, from Milky Way-like scales to large clusters, aligns more closely with observational data, marking a significant advancement in the predictive power of our semianalytic model.

In conclusion, echoing the statement of \cite{popesso2024}, an additional physical mechanism capable of expelling gas beyond the halo while preserving the cold gas and stellar components is essential for numerical methods. The hot gas ejection mode introduced in \texttt{FEGA25} appears to be a promising solution, effectively reducing the hot gas content within halos and improving agreement with observations. While there is still room for refinement, this mechanism provides a strong foundation. We plan to further explore its implications in an upcoming paper (Contini et al., in prep.), where we will examine the evolution of the baryon and hot gas fractions with redshift and disentangle the contributions of supernova and AGN feedback in expelling gas beyond the virial radius.

%%%%%%%%%%%%%%%%%%%%%%%%%%%%%%%%%%%%%%%%%%%%%%%%%%%%%%%%%%%%%%%%%%%%%%%%%%%%%%%%%%%%%%%%%%%%%%%%%%%%%%%%%%%%%%%%%%%%%%%%%%

\section*{Acknowledgements}
% The authors thank the referee XXXX for his very constructive comments which helped to improve the manuscript.
E.C. and S.K.Y. acknowledge support from the Korean National Research Foundation (RS-2025-00514475). E.C. and S.J. acknowledge support from the Korean National Research Foundation (RS-2023-00241934). All the authors are supported by the Korean National Research Foundation (RS-2022-NR070872). J.R. was supported by the KASI-Yonsei Postdoctoral Fellowship and by the Korea Astronomy and Space Science Institute under the R\&D program (Project No. 2023-1-830-00), supervised by the Ministry of Science and ICT.

%%%%%%%%%%%%%%%%%%%%%%%%%%%%%%%%%%%%%%%%%%%%%%%%%%%%%%%%%%%%%%%%%%%%%%%%%%%%%%%%%%%%%%%%%%%%%%%%%%%%%%%%%%%%%%%%%%%%%%%%%%
\section*{Data Availability}
Codes and galaxy catalogs used in this study can be obtained by contacting the corresponding author.

%%%%%%%%%%%%%%%%%%%%%%%%%%%%%%%%%%%%%%%%%%%%%%%%%%%%%%%%%%%%%%%%%%%%%%%%%%%%%%%%%%%%%%%%%%%%%%%%%%%%%%%%%%%%%%%%%%%%%%%%%%
\bibliography{paper}{}
\bibliographystyle{aasjournal}

\end{document}